\documentclass{aa}  

\usepackage{graphicx}
\usepackage{txfonts}
\usepackage{xcolor}
\usepackage{caption}
\usepackage{subcaption}
\usepackage{amssymb}
\usepackage{mathtools}
\usepackage{amsmath}
\usepackage{rotating}
\usepackage{hyperref}
\usepackage{cleveref}

\newcommand{\pc}{{\mathrm{pc}}}
\newcommand{\Myr}{{\mathrm{Myr}}}
\newcommand{\yr}{{\mathrm{yr}}}
\newcommand{\au}{{\mathrm{au}}}
\newcommand{\Msun}{{\mathrm{M}_\odot}}

\begin{document}

   \title{DAWN}

   \subtitle{I. Simulating the formation and early evolution of stellar clusters with Phantom N-Body}

   \author{Y. Bernard
          \inst{1}
          \and
          E. Moraux\inst{1}
          \and
          D. J. Price\inst{2}
          \and
          F. Motte\inst{1}
          \and
          F. Louvet\inst{1}
          \and
          I. Joncour\inst{1}
          }

   \institute{Univ. Grenoble Alpes, CNRS, IPAG, 38000 Grenoble, France
         \and School of Physics and Astronomy, Monash University, Vic. 3800, Australia
             }

   \date{Received April 1, 2025; accepted August 4, 2025}

  \abstract 
   {The star formation process produces hierarchical, clustered stellar distributions through gravoturbulent fragmentation of molecular clouds. Simulating stellar dynamics in such an environment is numerically challenging due to the strong coupling between young stars and their surrounding gas, and the large range of length and time scales.}
   {This paper is the first of a suite aimed at investigating the complex early stellar dynamics in star-forming regions, from the initial collapse of the molecular cloud to the phases of complete gas removal. We present a new simulation framework. This advanced framework is the key to generating a larger set of simulations, enabling statistical analysis, which is mandatory to address the stochastic nature of dynamical interactions.}
   {Methods originating from the stellar dynamics community, including regularisation and slowdown methods (SDAR), have been added to the hydrodynamical code Phantom to produce simulations of embedded cluster early dynamics. This is completed by a novel prescription of star formation to initialise stars with a low numerical cost, but in a way that is consistent with the gas distribution during the cloud collapse. Finally, a prescription for \ion{H}{II} region expansion has been added to model the gas removal.}
   {We have run testcase simulations following the dynamical evolution of stellar clusters from the cloud collapse to a few Myr. Our new numerical methods fulfil their function by speeding up the calculation. The N-body dynamics with our novel implementation never appear as a bottleneck that stalls the simulation before its completion. Our new star formation prescription avoids the need to sample individual star formations within the simulated molecular clouds with a large number of SPH particles. Overall, these new developments allow accurate hybrid simulations in minimal calculation time. 
   Our first simulations show that massive stars largely impact the star formation process and shape the dynamics of the resulting cluster. Depending on the position of these massive stars and the strength of their \ion{H}{II} regions, they can prematurely dismantle part of the cloud or trigger a second event of cloud collapse, preferentially forming low-mass stars. This leads to different stellar distributions for numerical simulations with similar initial conditions and confirms the need for statistical studies. Quantitatively, and despite the implementation of feedback effects, the final star formation efficiencies are too high compared with those measured in molecular clouds of the Milky Way. This is probably due to the lack of feedback mechanisms other than \ion{H}{II} regions, in particular jets, non-ionising radiation, or Galactic shear.}
    { Our new Phantom N-Body framework, coupled with the novel prescription of star formation, enables efficient simulation of the formation and evolution of star clusters. It enables the statistical analysis needed to establish a solid theoretical framework for the dynamical evolution of embedded star clusters, continuing the work done in the stellar dynamics community. 
   }


   \maketitle

\section{Introduction} \label{sec:Intro}

Star formation occurs within turbulent, collapsing molecular clouds in the Galaxy. The gravitational collapse of these turbulent clouds produces interconnected filaments and hubs that amplify density anisotropies. Observations with the \textit{Spitzer} telescope have revealed that the star formation process takes place preferentially inside these dense structures in star-forming regions \citep{evans_2009,guthermuth_2009, andre2014}. Denser regions with similar levels of turbulent and magnetic support are more sensitive to fragmentation, leading to a dominant `clustered' star formation channel. Consequently, the distribution of newborn stars is inherited from the distribution of filaments and hubs/ridges \citep{Schneider2010,Peretto2013,Motte2018a}, and appears to be structured, with multiple clusters embedded in their birth nebula/clump. 

Observations have shown that this clustered state is transient for $90 \%$ of stars \citep{lada_embedded_2003}, and that only $10\%$ of newborn stars remain in open clusters after $10~\Myr$. This dynamical evolution is fast and complex, and a better understanding of this evolution is needed to constrain environmental properties required in other fields. For example, the statistics of fly-by events during the early evolution of star-forming regions could help to constrain their effect on planet formation within protoplanetary disks \citep{cuello_2023}. 

Pure N-body simulations \citep{aarseth_hills_1972,goodwin_2004,bastian_2006,parker_dynamical_2014,wright_dynamical_2014} have tried to explain the rapid dissolution of natal star clusters with stellar dynamics. These simulations modelled young stellar clusters using an initially fractal spatial distribution, consistent with observational constraints \citep{cartwright_2004}. The results indicate that dynamical instabilities, such as close encounters or unstable triples, can lead to the complete dissolution of an initial cluster or to the conservation of a strongly bound cluster of stars at the end of the simulation. Furthermore, the statistical robustness of these findings is ensured by the averaging of hundreds of simulations to draw conclusions. However, pure N-body simulations ignore one key component of this dynamical evolution: the strong links and interactions between newborn stars and their gaseous environment. Previous N-body studies have attempted to account for the gas potential using simple radial analytical formulae with a decay in time to model the gas expelled from embedded clusters \citep{kroupa_2001}. More recent simulations have focused on the early phases of star formation, revealing the complex and time-variable nature of this coupling that influences both the dynamics and the formation of stars.\citep{verliat_influence_2022,Fujii_2021,grudic_starforge_2021,gonzalez_2020}. Hence, only direct methods can model star formation and stellar feedback simultaneously to follow the stellar dynamics during the early evolution of clusters in star-forming regions.

On the other hand, numerical studies focusing on star formation have naturally characterised this process using a direct hydrodynamic model. \cite{larson_numerical_1969} and \citet{boss_1979} focused on a single, isolated clump of gas gravitationally collapsing to form single or binary stars. Such works tried to self-consistently follow the collapse from the initial dense clump up to the onset of nuclear fusion \citep{bate_1998,bate_2014,wurster_2018}. Recent studies \citep[e.g.,][]{Ahmad_2024,Mauxion2024} continue to refine models of protostar formation, incorporating more realistic physical processes to predict phenomena such as the formation and evolution of protoplanetary disks that are essential to planet formation.

However, the star formation community rapidly understood that the environment is key to explaining stellar properties. Hence, it became critical to perform simulations at the molecular cloud scale while modelling the formation of individual protostars. \cite{bate_modelling_1995} developed a method called sink particles that hides dense parts of a collapsing clump inside a massive particle capable of accreting infalling material without computationally expensive hydrodynamic calculations. This method unlocked the ability to follow the evolution of star-forming regions beyond the point of initial collapse. As a consequence, sink particles are now used in almost every star formation simulation. 

Early studies \citep{bate_formation_2002,bate_origin_2005} simulated small, isolated molecular clouds with a diameter of a fraction of a parsec and an initial mass of tens of solar masses. Later developments enabled the modelling of radiative feedback from protostars on the surrounding gas, which was found to be essential to produce a more convincing stellar mass distribution  \citep[e.g.,][]{bate_importance_2009,Hennebelle2020}. Other studies tried to identify links between molecular cloud initial conditions, such as the temperature, magnetic field \citep{Myers2014,Lee_2019} or initial turbulent state \citep{Joos2013,liptai_does_2017}, and stellar properties like the mass or the spatial distribution. Most Recent studies have attempted to model physical processes in star-forming regions more realistically. Specifically, stellar feedback is thought to play a major role in both star formation and the global evolution of molecular clouds \citep{Krumholz_2014}. \cite{verliat_influence_2022} found that protostellar jets can regulate accretion around dense objects by injecting turbulent support inside the surrounding gas. State-of-the-art simulations attempt to model the evolution of giant molecular clouds through the entire star formation process, from the initial collapse to the gas removal mainly driven by \ion{H}{II} region expansion produced by massive stars \citep{guszejnov_cluster_2022}. For instance,\cite{wall_collisional_2019} and \cite{Fujii_2022} reused methods from the stellar dynamics community into their hydrodynamic codes to perform what is called a hybrid hydro/N-body simulation. 

These simulations aimed to bridge the gap between previous work by the stellar dynamics community and star cluster formation simulations, including stellar feedback. They provided a framework to explain the dynamical evolution of young star clusters influenced by their environment. However, they did not follow the statistical approach of previous pure N-body cluster simulations, which is needed to identify typical evolutionary tracks followed during the stochastic dynamics of stars in a cluster.

In this paper, we show how methods developed for accelerating N-body simulations and efficient star formation and stellar feedback prescriptions can be incorporated directly into a 3D hydrodynamics code, adding missing physics to pure N-body simulations. Such a framework allows us to initialise young embedded clusters directly from star formation in a molecular cloud collapse, leading to self-consistent cluster initial conditions linked with the natal gaseous environment. Their evolution can therefore be followed through the direct influence of the gas, itself influenced by gravity and stellar feedback. We aim to follow this evolution up to the emergence of clusters from their natal cloud. Our new framework has been designed to be fast enough to enable the statistical approach necessary to study the dynamical evolution of embedded clusters.

The paper is organised as follows. First $N$-body algorithms newly implemented in the code Phantom \citep{price_phantom_2018} are presented in Sect.~\ref{sec:nbody}. \Cref{sec:phys} describes the new star formation prescription (Sect.~\ref{sec:starform}), the implementation of \ion{H}{II} region feedback (Sect.~\ref{sec:HII}), and the fiducial model studied in this work (Sect.~\ref{sec:fidu}). Simulation results of this model are described and discussed in Sect.~\ref{sec:results}. \Cref{sec:discussion} discusses numerical performance and accuracy. Finally, \Cref{sec:conclusion} gives our conclusions and summarises future prospects. Paper II will then be dedicated to a deeper investigation of our new star formation prescription, focusing on the mass distribution and multiplicity statistics.

\section{Phantom} and N-body algorithms\label{sec:nbody}
Simulations presented in this work were performed with the 3D Smoothed Particle Magnetohydrodynamics (SPMHD) code Phantom \citep{price_phantom_2018}. Smoothed particle hydrodynamics \citep[SPH;][]{Lucy_1977,Gingold_1977} solvers use a Lagrangian approach to solve fluid equations by discretising the medium into particles of fixed mass that follow the fluid motion. Fluid forces are derived by calculating quantities on these particles with neighbour weighted sums using smoothing kernels \citep{price_2012}. We chose Phantom because of its versatility. It is capable of modelling a variety of astrophysical systems with its numerous physics modules, like an MHD solver (ideal and non-ideal), self-gravity, dust, chemical networks, sink particles and more. However, this code was not ready yet to perform embedded cluster simulations at the beginning of this work. While a few studies on star formation have already been carried out \citep{liptai_does_2017,wurster_gas_2023}, these studies stopped the evolution of their star-forming region at around the initial free fall time of the cloud $\tau_\mathrm{ff}$. This end time is limited by the large computational cost of such models. One of the growing constraints is the gravitational dynamics between newly formed stars when no softening is applied, which can then form in close binaries or multiple systems. Stars in such systems are subject to strong interactions that result in small time-step constraints that can dramatically slow down a simulation. 

In Phantom, gravitational forces between point masses (as stars) are computed using a direct method. This direct algorithm is coupled to a Leapfrog symplectic integrator to follow the dynamics of stars. This integrator is only 2nd order, which needs short timesteps to reach sufficient accuracy. This combination of methods is inefficient and can be inaccurate during the early evolution of embedded clusters. As the number of steps to integrate a cluster up to a few million years is prohibitive, it also implies an accumulation of truncation errors that can drastically change the evolution of close binaries or multiples, which are key in the overall cluster evolution. A hard binary disruption due to inaccuracy in the integration scheme can lead to the artificial fast expansion of the host cluster. These numerical effects motivate new implementations written in Phantom that reuse methods developed in collisional N-body codes. The SDAR method proposed by \cite{wang_slow-down_2020} and the forward symplectic integrator (FSI) recently presented in \cite{rantala_frost_2021} and \cite{dehnen_symplectic_2017}. For the sake of completeness, we describe our implementation from these original papers in App.~\ref{App:FSI}.

\subsection{Forward Symplectic integrator: FSI}
Symplectic integrators preserve conservation properties. They can conserve angular momentum to round-off error while constraining the energy error inside a small range, depending on the truncation order and the timestep choice. They achieve that by rewriting the time evolution operator. This new formulation implies that the integrated Hamiltonian is slightly different from the true one by what is called the error Hamiltonian. The art of building high-order symplectic integrators lies in minimising this error term \citep{yoshida_construction_1990}. The Forward Symplectic Integrator (or FSI) \citep{dehnen_symplectic_2017,rantala_frost_2021} achieves this minimisation up to fourth order with the unique property of producing no backwards substeps inside its scheme, which is essential if dissipative forces are in the scheme. More details on how to construct symplectic integrators are presented in Appendix \ref{App:FSI}. That aside, one can construct the new time evolution operator derived from this special method using
\begin{equation}
    e^{\delta t \, H} = e^{\delta t \, (T+U+\frac{\delta t^2}{48}G)} \approx e^{\frac{1}{6}\delta t \, U } e^{\frac{1}{2}\delta t \, T} e^{\frac{2}{3}\delta t \,(U+\frac{\delta t^2}{48}G) } e^{\frac{1}{2}\delta t \, T } e^{\frac{1}{6}\delta t \, U},
\end{equation}
 where $H$ is the Hamiltonian, $T$ the kinetic energy, $U$ the potential and $G$ an extra potential. All the coefficients in this time evolution operator are strictly positive. The middle operation is modified with what is commonly called a gradient force evaluation produced by a slightly different potential $U+\frac{\delta t^2}{48}G$ \citep[cf.,][]{rantala_frost_2021}. Knowing the definition of $G$ and for a system composed only of $N$ point masses coupled by gravitational interactions where $\mathbf{a_j}$ is the acceleration sensed by each mass, the gradient acceleration can be written as
\begin{equation}
    \mathbf{g}_i = -\frac{1}{m_i}\frac{\partial G}{\partial \mathbf{r}_i}  =\frac{\partial}{\partial\mathbf{r}_i}\sum_j^N\|\mathbf{a}_j\|.
    \label{gradientforce}
\end{equation}
Expanding the partial derivative in this equation leads to the common formula of the jerk computed in Hermite fourth-order integrator \citep{makino_hermite_1992} and gives
\begin{equation}
    \mathbf{g}_i =2\sum^N_{i = 1}\sum^N_{i\neq j} \frac{Gm_j}{r_{ij}}\left[r_{ij}^2\mathbf{a}_{ij}-3(\mathbf{r}_{ij}\cdot\mathbf{a}_{ij})\mathbf{r}_{ij} \right],
    \label{gradientforce2}
\end{equation}
where $\mathbf{a}_{ij}$ and $\mathbf{r}_{ij}$ are the relative acceleration and position between  particles $i$ and $j$. The modified acceleration of the middle kick operation for a particle $i$ is then
\begin{equation}
\mathbf{a}_{i*}=\mathbf{a}_i + \frac{\delta t}{48}\mathbf{g}_i.
\label{gravgradforce}
\end{equation}
This formula is not trivial to compute. First, all pairwise accelerations are required to compute each gradient term, which means that they need to be evaluated beforehand. During the middle kick phase, there is therefore one acceleration and one gradient acceleration evaluation consecutively to compute the final acceleration to kick the velocity of each body in the system. This method turned out to be expensive to implement in the current Phantom framework. Indeed, gradient accelerations in a simulation where softening is applied between stars and gas need to take this softening into account, which is possible but numerically complex. Instead, \cite{omelyan_extrapolated_2006} proposed to compute this modified acceleration by extrapolating it from a slightly shifted position of each body, which depends on the standard acceleration. Formally, it is written as
\begin{equation}
    \mathbf{a}_{i*} = \mathit{a}(\mathbf{r}_{i*}); \hspace{1cm} \textrm{where} \hspace{0.5cm} \mathbf{r}_{i*}\equiv \mathbf{r}_i+\frac{\delta t^2}{24}\mathit{a}(\mathbf{r}_{i})
    \label{gradextrap}
\end{equation}
and $\mathit{a}$ is the function to compute the acceleration for each particle in the system. One can see in this formula that there are still two acceleration evaluations. However, it reuses the same function in both cases, with a shifted position in the latter one. \cite{omelyan_extrapolated_2006} and \cite{chin_forward_2005} showed that this new formula gives results close to the original. This method is far easier to implement in an existing framework. Hence, we chose this method instead of the original one.

In terms of precision, FSI outperforms all other standard fourth-order symplectic schemes like Forest-Ruth integrator \citep{forest_fourth-order_1990} and even improved ones like Position extended Forest Ruth like integrator (PEFRL)\citep{omelyan_optimized_2002} (See Appendix \ref{App:FSI}). \cite{chin_forward_2005} argued that even with the same order of truncation error, the forward scheme will always outperform standard schemes because others evaluate forces far from the true trajectory of the object. By contrast, forward schemes evaluate forces very close to the true trajectory with the same time step resolution. In the end, standard schemes need smaller time steps to achieve the same resolution. FSI seems to be the best option for a new integration algorithm of point masses in Phantom. It has one of the best precisions for fourth-order schemes with limited supplementary force evaluation in one global step. Its forward nature assures that nothing can go wrong if dissipation is present in the system, which is often the case in astrophysical hydrodynamical simulations.

\subsection{Slow down Algorithmic regularisation: SDAR}
\subsubsection{Regularised integration}
The FSI presented above is an accurate integrator that already outperforms second-order leapfrog schemes. It is then possible to limit the number of steps per orbit needed to sample a good trajectory of each body. Typically, this number of sample points is set from 300 to 500 with the previous Leapfrog integrator. With the new FSI, we can easily divide this number by a factor of four while retaining better accuracy. 

However, updating the main integrator is not enough to resolve the stellar dynamics within an embedded cluster in a reasonable computation time. Stellar densities are high in cluster environments, and binaries and multiple systems represent a significant fraction of the population \citep{offner_origin_2023}. These systems can produce strong computational constraints during close encounters or with short-period binaries. For example, a binary with 10 $\au$ separation and a total mass of 2 $\text{M}_\odot$ has a period around $ 20 \, \text{yr}$. It could then produce a time step constraint around $\delta t = 0.33 \, \text{yr}$ when the typical free fall time of a molecular cloud in this study is around $5.5 \, \Myr$, which gives a large dynamic range. This simple example does not capture the stringent time step constraints that could appear in a young cluster. Day orbit period binaries and high eccentric ones make direct computation infeasible. 

The stellar dynamics community has already tackled these high dynamical scale problems to model the long-term evolution of stellar clusters containing a few thousand bodies to a few tenths of a million. To meet this challenge, sophisticated methods have been developed to manage the two main issues at large dynamic ranges, which are the gravitational singularity during pair interaction and the secular evolution of multiple systems.

First, gravitational pair interaction is a function of the inverse square distance between two bodies. It produces a singularity at the distance $d = 0$. This singularity can strongly constrain the simulation time step. During close encounters or periapsis approaches in a highly eccentric binary, the velocity of the two bodies can dramatically increase due to the inverse square distance proportionality of their interactions. Their true trajectory will be missed if there are not enough sample points. Worse, the pair interaction can produce false runaway bodies. In collisionless models as cosmological ones like Gadget 4 \citep{springel_simulating_2021}, the interaction is softened in a certain radius around each point mass to avoid these strong pair interactions. However, softening methods are prohibited in collisional N-body codes, as the collisional dynamics govern the future evolution of a cluster. The immediate solution could be to increase the number of steps. However, this would also produce large truncation and round-off errors due to the accumulation of steps in addition to being computationally expensive.

Various methods have been proposed in the literature for fifty years now \citep{kustaanheimo_perturbation_1965,mikkola_time-transformed_2002} to manage these two issues. All of them are based on the same idea: regularisation of gravitational interactions. The idea is to hide the singularity. One manipulates the equations of motion by applying spatial and time transformations on the system of equations to make the ${1}/{d^2}$ term in the force evaluation vanish. The first regularisation method used in this field was proposed by \cite{kustaanheimo_perturbation_1965} (KS method). It transforms binary gravitational motion into a simple harmonic oscillator, meaning that the time step can be constant during a whole period of the binary, whatever its eccentricity. The details of this transformation can be found in the original paper and in \cite{aarseth_regularization_2003}. 
It has been successfully used in several collisional codes in the past, such as the series of Aarseth's N-Body codes \citep{aarseth_nbody1_1999}. However, its complexity makes it hard to implement. In addition, KS regularisation of multiple systems requires specialised methods, which add up to a rather complex method. More recent codes \citep{wang_petar_2020,rantala_bifrost_2023} preferred a different approach to reach the same results, using what is called an \emph{algorithmic regularisation} (AR) \citep{mikkola_algorithmic_1999,preto_class_1999}. This work follows the same approach.
AR methods apply a time transformation to the system
\begin{equation}
    \mathrm{d}s = \Omega(\mathbf{r})\mathrm{d}t \label{eq:ttrans},
\end{equation} 
where $s$ is a new time integration variable that is linked with the real time $t$ by a time transformation function $\Omega(\mathbf{r})$. This function is often taken as the gravitational potential of the system.

By contrast to KS, it leaves the spatial coordinates unchanged. The singularity is still present in a pair interaction. However, the idea is then to use a specialised algorithm to produce regular results. There are two main algorithms to achieve that. The LogH method presented in \cite{rantala_bifrost_2023} or \cite{aarseth_regularization_2003} and the time-transformed leapfrog (TTL) \citep{mikkola_time-transformed_2002}. We adopt the latter method in this work, which is described in detail in Appendix \ref{APP:TTL}. Briefly, it is possible to construct a new symplectic integrator using the modified system of equations. The change of variable produces a new integrated momentum-like variable $W$ defined as
\begin{equation}
    \frac{\mathrm{d}W}{\mathrm{d}s} = \sum_{i=1}^N \frac{1}{\Omega}\frac{{\rm d}\mathbf{r}_i}{{\rm d} t}\cdot \frac{\partial \Omega}{\partial \mathbf{r}_i}.
\end{equation}
This new variable will then have the same behaviour as velocities in a symplectic scheme. The Leapfrog method that lends its name to this method is not precise enough in such applications. Two solutions have been used to reach a satisfactory accuracy. First, the original algorithm from Mikkola's code and the new code BIFROST \citep{rantala_bifrost_2023} used the Gragg–Bulirsch–Stoer (GBS) extrapolation method \citep{gragg_extrapolation_1965,bulirsch_numerical_1966} to ensure a user-chosen relative error. The second solution used in PeTar \citep{wang_petar_2020}, another stellar dynamics code, is to integrate the system with a high-order symplectic scheme from \cite{yoshida_construction_1990}. We chose the latter method for its simplicity. However, GBS extrapolation will be tested in further work.

The essence of this method can be understood from the time transformation (Eq.~\ref{eq:ttrans}). The link between the new integration variable $s$ and the real time $t$ depends on the potential strength. This way, $s$ can have constant large time steps as the integrated time $t$ will scale depending on the potential strength. It means that when the gravitational forces are strong (close to the singularity), the time integration will be slow and algorithmically adjusted to the potential strength. That way, singularity approaches are not an issue anymore even if it is still present in equations.

\subsubsection{Slowed down binaries}

The last challenge to tackle is the secular evolution of perturbed binaries and multiple systems. Regularised integrators can solve collisional trajectory issues. Nevertheless, they still need a few sample points to integrate orbits. It means that very hard, eccentric, and weakly perturbed binaries or stable triple systems can bottleneck the computation, as the dynamic time range between a cluster and these systems can be extremely large. Because of their hardness, these binaries can be considered unperturbed, and thus their motion can be found analytically using Kepler solvers. It resolves only a part of the issue, as stable triples are commonly found in simulations and close encounters in dense stellar regions can greatly perturb binary systems. \cite{mikkola_slow-down_1996} proposed an elegant solution with the cost of losing the orbital phase of these hard binaries. The slowdown method proposes to slightly modify the original Hamiltonian of a weakly perturbed binary according to
\begin{equation}
    H = \frac{1}{\kappa}H_b+(H-H_b),
\end{equation}
where $\kappa$ is called the slow down factor and $H_b$ is the unperturbed binary Hamiltonian. This equation means that only the binary dynamics are scaled down by a factor $\kappa$. Original velocities are conserved by this scaling. Equations of motion derived from this Hamiltonian are given by
\begin{eqnarray}
    \frac{\mathrm{d}\mathbf{v}_i}{\mathrm{d}t} &=& -\frac{Gm_c(\mathbf{r}_i-\mathbf{r}_c)}{\kappa\|\mathbf{r}_i-\mathbf{r}_c\|^3} -\frac{\partial U(\mathbf{r}_i)}{\partial\mathbf{r}_i},\label{eq:rSD}\\
    \frac{\mathrm{d}\mathbf{r}_i}{\mathrm{d}t} &=& \frac{\mathbf{v}_i}{\kappa},\label{eq:vSD}
\end{eqnarray}
where $U$ is an arbitrary external potential and $i$ and $c$ index points on one binary member and its companion, respectively. One can see that the forces due to gravitational interactions between the primary and the secondary are scaled down by $\kappa$. The original orbital period of the binary is divided by $\kappa$. It is then straightforward to understand that the time constraint on the binary will then be relaxed by the same factor. 

Equations \ref{eq:rSD} and \ref{eq:vSD} show that for a given binary system, its orbital parameters are conserved by the scheme, despite perturbations that will be discussed shortly. It also means that even if the apparent period is $\kappa$ times longer than the original one, the true orbital parameters are conserved by the scheme. The only change comes from the external perturbation.

\cite{wang_slow-down_2020} showed that applying this method on one slowed-down period $P_\mathrm{sd} = \kappa P$ gives the same velocity modification as if the system was integrated during $\kappa P$. It means that the slowdown method averages perturbations applied to the binary on a slowed-down period $P_\mathrm{sd}$. Nonetheless, it is only applicable if the perturbation is weak enough. Thus, it means that we should choose $\kappa$ carefully to preserve the accuracy of the averaging. On intermediate perturbed binaries, this averaging property is relevant only for a few orbits. By contrast, external perturbations can be almost negligible for a hard binary. In that case, the averaging property can be accurate enough on several orders of magnitude of the original orbital period. 

The generalisation to a few-body problem is straightforward and detailed in \cite{wang_slow-down_2020}. Only two changes are needed: i) the bulk motion of a whole binary in an $N$-body system is not slowed down and ii) the perturbation on a binary of the system is now the summation of each direct gravitational force produced by each perturber inside the system. As shown above, $\kappa$ quantifies the number of original periods where the perturbation applied to the binary can be averaged. On a highly perturbed binary, it is impossible to average the perturbation on multiple periods, which means $\kappa = 1$. A hard binary can stay unperturbed during a whole simulation. The averaging is not risky at all, and then $\kappa$ can stay at a high value. However, one can see that this factor needs to scale with the strength of the perturbation. A special method has been added to our implementation to modulate $\kappa$ depending on external perturbations following \cite{mikkola_slow-down_1996}. A detailed overview is presented in \Cref{APP:SD}, and a unit test performed to validate the implementation is presented in \Cref{App:SDARtest}.

The Slow Down method can easily be mixed up with Algorithmic regularisation, which then gives what is called a SDAR integration method \citep{wang_slow-down_2020}. This method gives an efficient integrator for multiple systems, which are common in young stellar clusters. It can be used in combination with standard integrators such as the FSI presented above to form a complete stellar dynamics framework capable of evolving an entire cluster of stars for a long time.

\subsection{Phantom new N-body Implementation}

We implemented the integration methods described above into the public version of Phantom. They could not fit directly into the main integration routine without a major refactor. Importantly, the SDAR method is used to integrate the motion of stars but not SPH gas particles. Thus, the new methods are applied to a specific part of the system and not the entire one.

The original Phantom integration scheme was derived from the reference system propagator algorithms (RESPA) \citep{tuckerman_reversible_1992} which split forces applied on particles in the system into two categories, the ``slow'' and ``fast'' forces with respectively weak and strong time step constraints. This splitting allows the construction of a two-level leapfrog scheme according to
\begin{eqnarray}
    &\mathbf{v} = \mathbf{v} + \frac{1}{2}\Delta t_{\text{sph}} \mathbf{a}_{\text{sph}}\\
    &\text{N substeps}\left\{
    \begin{array}{l}
    \mathbf{v} = \mathbf{v} + \frac{1}{2}\Delta t_{\text{fast}} \mathbf{a}_{\text{fast}} \\

    \mathbf{r} = \mathbf{r} + \Delta t_{\text{fast}} \mathbf{v} \\

    \mathbf{a}_{\text{fast}} = \mathbf{a}_{\text{fast}}(\mathbf{r}) \\

    \mathbf{v} = \mathbf{v} + \frac{1}{2} \Delta t_{\text{fast}} \mathbf{a}_{\text{fast}}\\

    \end{array}
    \right.\\
    &\mathbf{a}_{\text{sph}} = \mathbf{a}_{\text{sph}}(\mathbf{r})\\
    &\mathbf{v} = \mathbf{v} + \frac{1}{2} \Delta t_{\text{sph}} \mathbf{a}_{\text{sph}},
\end{eqnarray}
where $\mathbf{a}_{\text{sph}}$ corresponds to SPH accelerations (pressure, viscosity, self-gravity...). $\Delta t_{\text{sph}}$ is the time step associated with them and used in the first leapfrog level. The drift phase of the first-level leapfrog is replaced by multiple substeps of another leapfrog in association with a smaller time step $\Delta t_{\text{fast}}$. This sub-timestep is controlled by faster accelerations contained in $\mathbf{a}_{\text{fast}}$ such as external potential, sink-gas, sink-sink accelerations, etc... 

The methods we presented above relate to stars, which correspond to point masses in Phantom (sink particles without softening). It means that only the second level needs to be updated, as it is where sink particles are updated during integration. The first upgrade was to switch the coarse Leapfrog integrator for the more accurate FSI presented above. The substep integrator can then be rewritten as
\begin{eqnarray}
    \mathbf{v} &=& \mathbf{v} + \frac{1}{6}\Delta t_{\text{fast}} \mathbf{a}_{\text{fast}}\\
    \mathbf{r} &=& \mathbf{r} + \frac{1}{2}\Delta t_{\text{fast}} \mathbf{v}\\
    \tilde{\mathbf{a}}_{\text{fast}} &=& \mathbf{a}_{\text{fast}}(\mathbf{r}_*); \hspace{1cm}\mathbf{r}_* = \mathbf{r}+\frac{\Delta t_{\text{fast}}^2}{24}\mathbf{a}_{\text{fast}}(\mathbf{r})\\
    \mathbf{v} &=& \mathbf{v} + \frac{2}{3}\Delta t_{\text{fast}} \tilde{\mathbf{a}}_{\text{fast}} \\
    \mathbf{r} &=& \mathbf{r} + \frac{1}{2}\Delta t_{\text{fast}} \mathbf{v}\\
    \mathbf{a}_{\text{fast}} &=& \mathbf{a}_{\text{fast}}(\mathbf{r}) \\
    \mathbf{v} &=& \mathbf{v} + \frac{1}{6} \Delta t_{\text{fast}} \mathbf{a}_{\text{fast}},
\end{eqnarray}
where $\tilde{\mathbf{a}}_{\text{fast}}$ is the extrapolated gradient acceleration needed to properly cancel 2nd order error terms in the FSI. We compute this gradient acceleration by evaluating forces twice in a row with a slightly shifted position during the second evaluation.

The SDAR method is not meant to integrate the motion of an entire cluster. It is specialised in few body evolution. Detection of subgroups of close stars is therefore mandatory to apply the SDAR method only on small groups. Gas particles are excluded from such integration methods as their gravitational softening and hydrodynamic forces can introduce unexpected behaviours in the integration. The group finding algorithm is directly derived from the one proposed in \cite{rantala_bifrost_2023}. Star neighbours are identified by searching them within a fixed searching radius $r_{\text{neigh}}$ around each star. The typical value used in this work was equal to 0.001 pc. A special criterion allows distant fly-by stars to enter a group if they enter the searching radius $r_{\text{neigh}}$ at the next step. It gives
\begin{equation}
    r \geq r_{\text{neigh}} \quad \textrm{and} \quad C\frac{r}{v}< \Delta t_{\text{fast}},
\end{equation}
where $r$ and $v$ are the relative position and velocity between the two stars tested, $C$ is a safety factor usually set to 0.1. This neighbour search is then used to construct the adjacency matrix of the entire system of stars in the simulation. Subgroups are then identified by searching connected components associated with this matrix using a depth-first search algorithm. On-the-fly binary detection is applied to each newly formed subgroup, which is essential for the slowdown method. This identification is done every substep to update subgroups as the system evolves. 

Subgroups can then be integrated with SDAR methods separately from the entire system during the FSI drift phase. However, as explained in \cite{rantala_bifrost_2023}, each star in subgroups still receives perturbations (from gas and point masses in Phantom) from their environment during the FSI kick phase. It then gives the full new integration scheme in Phantom given by
\begin{eqnarray}
    &\mathbf{v} = \mathbf{v} + \frac{1}{2}\Delta t_{\text{sph}} \mathbf{a}_{\text{sph}}\\
    &\text{N substeps}\left\{
    \begin{array}{l}
    \mathbf{v} = \mathbf{v} + \frac{1}{6}\Delta t_{\text{fast}} \mathbf{a}_{\text{fast}}\\
    \mathbf{r} = \mathbf{r} + \frac{1}{2}\Delta t_{\text{fast}} \mathbf{v}\\
    \mathbf{r}_{\text{sub}} = \text{SDAR}(\mathbf{r}_{\text{sub}},\mathbf{v}_{\text{sub}})\\
    \tilde{\mathbf{a}}_{\text{fast}} = \mathbf{a}_{\text{fast}}(\mathbf{r}_*); \hspace{1cm} \mathbf{r}_* = \mathbf{r}+\frac{\Delta t_{\text{fast}}^2}{24}\mathbf{a}_{\text{fast}}(\mathbf{r}) \\
    \mathbf{v} = \mathbf{v} + \frac{2}{3}\Delta t_{\text{fast}} \tilde{\mathbf{a}}_{\text{fast}} \\
    \mathbf{r}_{\text{sub}} = \text{SDAR}(\mathbf{r}_{\text{sub}},\mathbf{v}_{\text{sub}})\\
    \mathbf{r} = \mathbf{r} + \frac{1}{2}\Delta t_{\text{fast}} \mathbf{v}\\
    \mathbf{a}_{\text{fast}} = \mathbf{a}_{\text{fast}}(\mathbf{r}) \\
    \mathbf{v} = \mathbf{v} + \frac{1}{6} \Delta t_{\text{fast}} \mathbf{a}_{\text{fast}} \\
    \end{array}
    \right.\\
    &\mathbf{a}_{\text{sph}} = \mathbf{a}_{\text{sph}}(\mathbf{r})\\
    &\mathbf{v} = \mathbf{v} + \frac{1}{2} \Delta t_{\text{sph}} \mathbf{a}_{\text{sph}},
\end{eqnarray}
As the original Phantom integration scheme, the first integration level uses an individual time step scheme. However, the substep integration sticks with a global time step $\Delta t_{\text{fast}}$ chosen to be the most stringent constraint induced by fast forces. Subgroups only evolve during the SDAR phase and are excluded from the standard FSI drift routine. The SDAR integration routine uses the TTL regularisation method in association with a 6th or 8th-order Leapfrog scheme derived from \cite{yoshida_construction_1990} following \cite{wang_slow-down_2020}. On top of that, a slowdown is applied to binaries inside subgroups. For a group composed of only one binary, the slowdown factor $\kappa$ is computed using outside perturbations and is constant during the whole SDAR integration. For multiple system groups, $\kappa$ is computed with outside perturbations and perturbations induced by other group members. This $\kappa$ needs to be updated at each SDAR step as inside perturbations evolve following the group dynamics. A correction of the new variable $W$ introduced in the TTL needs to be applied at each kappa update to conserve all system properties. This correction has the form of
\begin{equation}
    W = W + (\Omega(\kappa(t+dt)) - \Omega(\kappa(t))),
\end{equation}
where $\Omega$ is the gravitational energy of the subgroup.

\section{Physical models} \label{sec:phys}
\subsection{Star formation prescription}\label{sec:starform}
In order to save computational time and enable the statistical analysis of the evolution of embedded clusters, we do not follow the molecular cloud collapse all the way to the formation of stars. Instead, we adopt a low numerical resolution and use a star formation scheme below this scale. This section provides a detailed description of this model

\subsubsection{Sink particles and resolution criteria}

The star formation scheme is based on dynamic sink particle creation, already implemented in Phantom. It uses the criteria proposed in \cite{bate_modelling_1995}. Collapsing clumps that reach a user-defined threshold density $\rho_c$ will be eligible to pass a series of tests to verify that this clump will indeed quickly collapse. For example, a clump needs to be at least in marginal virial equilibrium. Formally, this gives
\begin{equation}
    \alpha_j+\beta_\text{rot} \leq 1,
\end{equation}
where $\alpha_j$ and $\beta_\text{rot}$ are the ratios of thermal and rotational energy over the gravitational energy of the clump, respectively. If the tests pass, the clump is then directly accreted onto a massive particle called a sink particle created at the centre of mass of the clump. The sink particle has an associated volume defined by a user-chosen accretion radius $r_\text{acc}$, within which gas particles are directly accreted. 

An essential point that has long been debated in the literature \citep{bate_resolution_1997,bate_formation_2002,hubber_resolution_2006} is the choice of this accretion radius together with the critical density value. The latter is typically chosen to resolve the initial Jeans mass of the clump to avoid numerical artefacts that could artificially fragment the collapsing clump. Thus, the maximal $\rho_c$ and minimal $r_\text{acc}$ are linked and depend on the mass resolution of the simulation.
Following \cite{hubber_resolution_2006}, this link is expressed as
\begin{equation}
    N_{\text{neigh}}\,m \leq M_{\text{Jeans}},
    \label{jeans_res}
\end{equation}
where $N_\text{neigh}$ is the constant number of neighbours of a particle in an adaptive smoothing length framework (around 60 in Phantom) and $m$ is the mass of one SPH particle. Therefore, the quantity $N_\text{neigh}m$ gives the minimum mass contained in an SPH kernel surrounding one particle, thereby corresponding to the minimum resolved clump mass in the simulation. As mentioned above, a clump that can turn into a sink needs to be resolved, i.e., $M_\text{clump} = M_{\text{jeans, min}}$. Using equation (\ref{jeans_res}) and the expression of the Jeans length
\begin{equation}
    \lambda_{\text{Jeans}} = \sqrt{\frac{\pi c_s^2}{G\rho}},
\end{equation}
One can find the maximum critical density:
\begin{equation}
    \rho_{c,\text{max}} = \left(\frac{\pi c_s^2}{G}\right)^3 \left(\frac{\pi}{6N_{\text{neigh}}m}\right)^2.
    \label{eq:rhomax}
\end{equation}
This maximum value can then be used to find the minimum accretion radius. Eq.~(\ref{jeans_res}) can be rewritten using the Jeans length as
\begin{equation}
    \lambda_{\text{Jeans}} \geq 4h,
\end{equation}
where $h$ is the smoothing length. Here, the number $4$ is derived from the size of the SPH kernel support which is equal to $2$ for the default cubic spline kernel in Phantom. The minimal accretion radius $r_{{\text{acc, min}}}$ linked with $\rho_{c_{\text{max}}}$ can then be expressed as
\begin{equation}
    r_{{\text{acc, min}}} = \frac{1}{2}\sqrt{\frac{\pi c_s^2}{G\rho_{c,\text{max}}}}.
    \label{eq:raccrho}
\end{equation}

Sink particles are often used to model one single star in formation \citep{bate_formation_2002,lee_stellar_2018}, which implies resolving the opacity limit for fragmentation around a few Jupiter masses \citep{Low_1976}. This can only be done for relatively small initial molecular clouds to keep the computational time reasonable. For example, the initial mass of the cloud in \cite{bate_origin_2005} was set to $50~\Msun$ using 3.5 million SPH particles, allowing them to resolve the minimum Jeans mass in their simulations ($0.0011~M_\odot$) with about 80 SPH particles. The sink particle accretion radius was correspondingly set to 5~$\au$, which is the typical size of the first Larson core \citep{larson_numerical_1969}. It ensures that only one star will form in a sink. Following star formation down to this scale within an entire giant molecular cloud would require a tremendous number of SPH particles and is currently beyond the reach of CPU-based codes like Phantom.

Other methods have been proposed to describe star formation within GMC following what was done in galaxy-scale simulations \citep{Renaud_2017}. In these simulations, sink particles are sufficiently large to represent small clusters of stars that form together in the same volume. A similar approach recently developed in \cite{wall_collisional_2019} considers sink particles as a stellar forge that produces stars from a list of masses sampled at the birth of each sink. As gas accretes onto these sinks, stars are generated at sink positions following the list of masses. \cite{liow_grouped_2022} proposed to group several sinks on the same list to produce more massive stars that cannot be generated by just one.

However, these methods introduce unwanted behaviour. For example, when stars are generated in a sink, the sink remains in the simulation with almost no mass in its reservoir. When accretion restarts, the sink position and velocity are mass-averaged with new accreted gas. As the sink mass is near zero, its position jumps to the accreted gas position. Due to the number of stars generated, sink particles in these simulations exhibited artificial Brownian-like motions. The erratic motion of these sinks can completely change the gas infall geometry and then perturb the spatial distribution of stars in the cluster.

\subsubsection{New star formation prescription from sink particles}
We propose a new prescription to model star formation using ``large'' sink particles where the fragmentation and accretion processes are not yet over. Typically, accretion radii are set to a few thousand $\au$ in this prescription. In that sense, sinks can be compared to the dense cores observed in star-forming regions.

However, these sinks are not necessarily cores from an observational point of view. They start their life as collapsing or marginally stable clumps. However, their future evolution is based on the sub-sink scale physics that will be described in the following. This evolution can diverge from what is called a core in observations. \cite{sadavoy_embedded_2017} studied YSOs within hot cores of a radius roughly equal to $10\,000$ $\au$ in the Perseus molecular cloud. They concluded that only a few (up to 5) YSOs formed in the surveyed cores where they saw multiples (24 embedded cores). These results are used as the main hypothesis for this prescription. A sink particle is considered as a ``core'' of a few thousand $\au$ where the fragmentation process will occur to form several (up to 5) YSOs. During a specified time the sink will continue to accrete infalling material that will feed the star ``seeds'' inside it. Nevertheless, physical processes inside a sink are not computed in the simulation (star seeds dynamics and accretion). 

The accretion time is hard to constrain, especially as a wide range of processes, such as outflows, dynamical encounters and radiation, influence it. These processes depend on the local environment and history of each core. Hence, it is very hard to establish a canonical value. \citet{dunham_young_2015} estimated the accretion time for pre- and protostellar phases with differences of a few orders of magnitude from $5\times10^4$ to a bit less than $1\times10^6$ years. These estimates are mainly for low mass objects; constraints on high mass cores are even looser. Here, we fix the accretion time $t_\text{acc}$ to an arbitrary median value of $0.5~\Myr$. This is sufficiently long to accumulate material to form high-mass stars, but sufficiently short to ensure the dynamical stability of star seeds inside the sink. 

Sinks can accrete during $t_\text{acc}$, but stellar objects inside them do not form immediately after sink formation because the fragmentation process inside a gas clump is not instantaneous compared to the accretion phase. Knowing if seeds are already created is essential in a merger of two sinks. A first guess for the delay time could be the free-fall time $\tau_\text{ff} = \sqrt{\frac{3\pi}{32G\rho}}$ of the clump just before the sink formation, where $\rho $ is the critical density for sink formation. In this work this value is around $1\times10^{-18} \mathrm{g}\,\mathrm{cm}^{-3}$ which gives a clump free fall time around $7\times10^{4}$~yr. The formation of gravitationally hard stellar cores may take longer. After reaching a certain density where heating and feedback occur, the infall is not free fall anymore and a slightly higher value could then be more accurate. We set the formation time of stellar objects in sinks $t_\mathrm{seed}\,$ to $0.1~\Myr$. This time is analogous with $t_\text{collapse}$ in \cite{Bleuler2014}.

\begin{figure}
    \centering
    \includegraphics[width=0.49\textwidth]{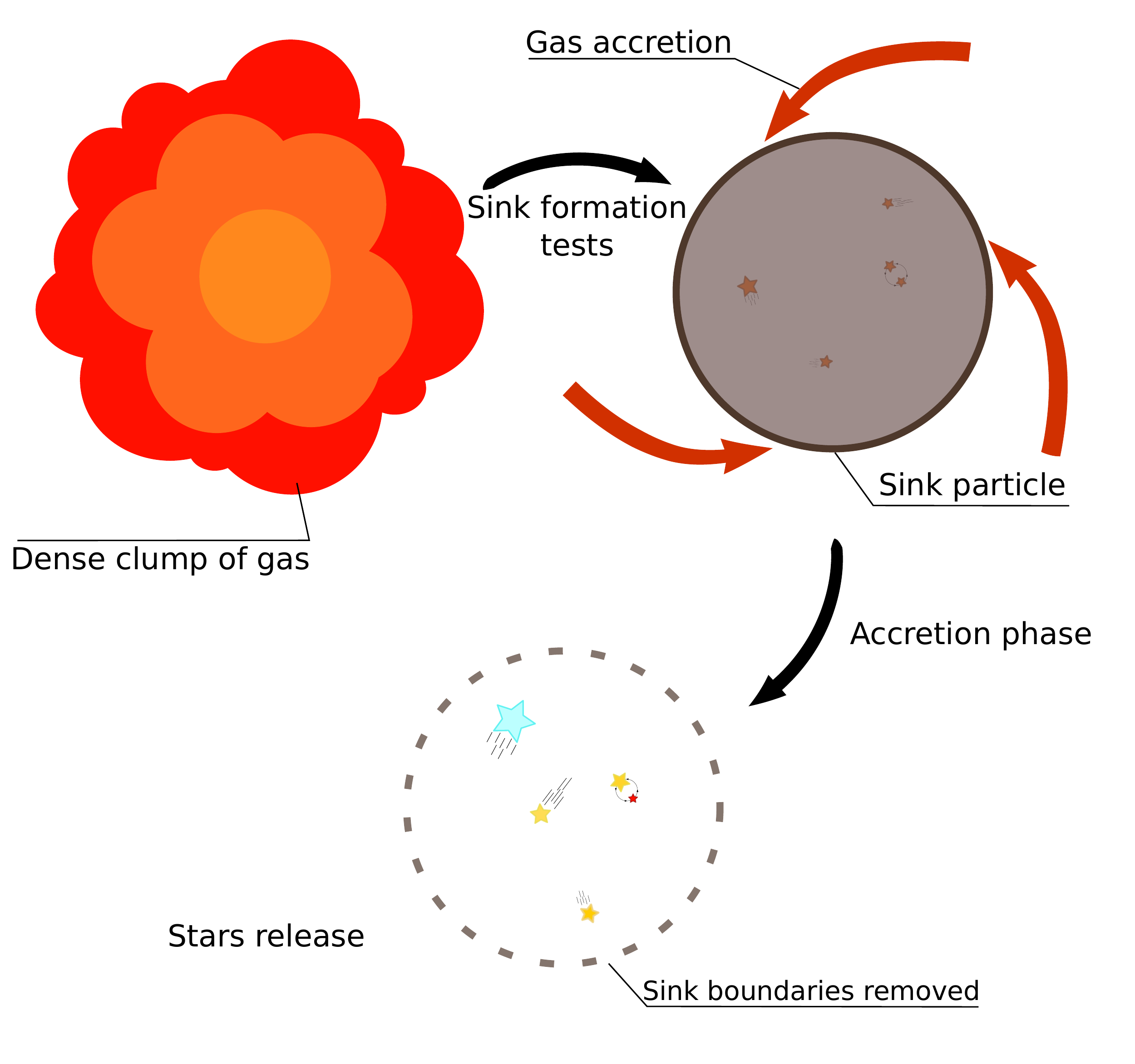}
    \caption{The star formation process described by our new prescription, from the clump transformation into sink particles to stars release inside the simulation domain. This process is subdivided into three parts. First, a collapsing clump of gas passes tests to be transformed into a sink particle. The sink particle can accrete during $t_\mathrm{acc}$. After reaching this time, the sink is dissolved and its mass is shared to produce newborn stars at its last location.}
    \label{fig:starform}
\end{figure}

Newborn stars are then released in the cluster in formation after a sink reaches its maximum accretion time. To summarise, this new star formation procedure using sink particles in Phantom is illustrated in Fig.~\ref{fig:starform} and proceeds as follows:
\begin{enumerate}
    \item First, a sink is formed when a clump of gas reaches certain conditions;
    \item The sink accretes surrounding materials continuously ;
    \item At the time $t_\mathrm{seed}$, the number of star seeds inside the sink is sampled randomly between 1 and 5. Accretion continues;
    \item When $t_\mathrm{acc}$ is reached, accretion is over. The sink is dissolved and young stars are released within its volume, keeping its momentum. Star masses are set by sharing randomly the parent sink mass reservoir. Positions and velocities are sampled to respect the dynamical stability of the group as described below;
    \item If sinks intersect each other by $r_\text{acc}$ during their lifetime, they merge to form one single sink. In that case, if star seeds have already formed, then their number will be summed up into the merged one. If not, only the mass and the momentum of both sinks are added to the merged sink.
\end{enumerate}

As said above, stars released after accretion onto their parental sink need to be distributed in their birth vicinity. Each new star will have a mass, position, and velocity that need to respect the main hypothesis of this prescription. There is conservation between sink mass and star mass. Therefore, sink mass is shared between child stars. The simplest way to share sink mass between stars is a random sharing with a cut at 0.08~$\Msun$ to ensure the formation of stars and avoid brown dwarf and substellar mass objects. We implement this by sampling $n_\text{seed}-1$ random numbers between zero and one. That way, a unit segment is subdivided into multiple sub-segments of random length, and their sum is always equal to one. Multiplying these segments by the mass of the sinks then produces a set of masses randomly chosen from the available mass in the reservoir. In addition to that, we resampled some points to ensure that no sub-segments are lower than 0.08~$\Msun$.

Finally, star velocities and positions should produce a bound system inside the sink volume. Many solutions exist to sample these quantities and the sampling method will greatly influence the resulting cluster morphology. The choice thus plays a key role in the simulation. However, exploring such a large parameter space would be computationally expensive and hard to implement on the first try. We opt for a simplistic approach using the same method that produces Plummer-like clusters in stellar dynamics-oriented simulations \citep{aarseth_comparison_1974}. It can produce spherical, centrally peaked, virialised star distributions with a very small development cost. After their formation, stars are not able to accrete material and are considered as point masses.

Stars and sink particles can coexist during the simulation. Their time integration is handled the same way in the code. We use the original method from \cite{Dehnen_read2011} as described in \cite{price_phantom_2018} to find the minimum time step constraint imposed by sink-sink, star-star or sink-star interactions. This one is used for the substep integration if it is the smallest of all forces computed during this phase. If these particles belong to a subgroup, then the integration is decoupled and handled by subgroup routines. To properly handle the collisional dynamics of stars, no softening is allowed in sink-sink, star-star or sink-star interactions. However, interactions between sink/star and SPH particles are softened as the latter cannot be considered as point-mass particles. The softening length $r_\text{soft-gas}$ is chosen to be the maximum of the gas smoothing length and the accretion radius $r_\text{acc}$ of sink particles (softening lengths are equal to $r_\text{soft}=4000\,\au$ for stars). Softening these forces avoids creating a large concentration of dense gas around stars like discs or infall, preventing a slowdown of the simulation. However, if a clump of gas reaches the condition to form a sink in the vicinity of a star, the sink is still allowed to form.

\subsection{Stellar feedback: \ion{H}{II} regions} \label{sec:HII}
Newborn stars inside active star-forming regions are connected with their natal environment. They inject both kinetic and thermal energy into their surroundings \citep[referred to as stellar feedback;][]{Krumholz_2014}, which can influence future star formation. Examples of momentum-driven feedback are jets or winds. Thermal feedback occurs from stellar irradiation.

Irradiation consists of two main components: ionising, and non-ionising radiation. Non-ionising radiation from direct stars or dust-reprocessed light can heat the gas via absorption. It occurs mostly at infrared wavelengths around dense cores hosting protostars, where the gas starts to be optically thick \citep{bate_importance_2009}. Ionising radiation is produced by massive stars ($>8~\Msun$) inside molecular clouds. Photons with energy larger than $E_\mathrm{Ly} = 13.2 \mathrm{eV}$ can directly ionise hydrogen atoms. Ionised gas around massive stars is heated to $10\,000~\mathrm{K}$ (the result of heating and cooling equilibrium). It creates a strong temperature gradient with cold neutral gas inside the region, which consequently produces a strong pressure gradient. Thus, ionising radiation from massive stars produces bubbles of hot gas around them that will expand inside the star-forming region, creating what is called an \ion{H}{II} region \citep{Osterbrock_2006}. This is thought to drive gas expulsion from star-forming regions, a key element of young stellar cluster evolution.

N-body simulations \citep{kroupa_2001} tried to model gas expulsion with simple methods like an exponential decay of a gravitational potential inside the star cluster. Since we model the gas directly, it is possible to model this feedback process by heating the gas around massive stars inside the newborn cluster. Models trying to reproduce \ion{H}{II} regions already exist and have been used in numerical simulations from the Galactic scale to the near-neighbourhood of massive stars. They can be modelled with more or less precision depending on trade-offs that authors made between accuracy and performance in agreement with their physical constraints. For example, \cite{Hopkins_2012} used a coarse model to implement \ion{H}{II} region feedback from entire star-forming regions to probe their effects on a global galaxy simulation without killing numerical performance. At the more sophisticated end, radiation transfer specialised codes are used in \cite{dale2012} and \cite{verliat_influence_2022} to precisely model ionising photon propagation inside a specific star-forming region to resolve the coupling between star formation and stellar feedback. However, such schemes come with a large numerical cost. Since our aim is to be computationally efficient while conserving individual star resolution, we want to model each \ion{H}{II} region produced by massive stars inside one simulation.

To achieve this, we implemented a model similar to that of  \cite{Fujii_2021} and \cite{gonzalez_2020} to simulate individual massive star \ion{H}{II} regions inside star-forming regions. This method is based on a distribution of the ionising photon rate produced by massive stars ($>8~\Msun$) at each SPH particle around them, assuming ionisation equilibrium inside the hot region and neglecting diffusion. If the ionising rate is sufficient to completely ionise a particle, the particle is tagged as ionised and its temperature and sound speed are changed accordingly. If not, it means that the ionisation front is nearby and all particles inside the \ion{H}{II} region have been tagged as ionised. 

Each massive star ($>8~\Msun$) ionising photon rate $\Delta N$ is computed using a fit function of the mass proposed in \cite{Fujii_2021} and based on observational surveys. Then, at each simulation update, an iterative algorithm verifies if the nearest SPH particles of a massive star are ionised or not. The algorithm computes the ionising photon rate to fully ionise the massive particle. Following \cite{Hopkins_2012}, it gives
\begin{equation}
    \delta N_i = \frac{m_i}{\mu m_p}\alpha_\beta n_i,
\end{equation}
where $m_i$ is the particle mass, $\mu$ the mean molecular weight, $m_p$, the proton mass, $\alpha_\beta\approx 2.3\times 10^{-13} \mathrm{cm}^3\mathrm{s}^{-1}$ the recombination rate, and $n_p$ the number density. If $\delta N_p<\Delta N$ the particle is ionised and $\delta N_p$ is subtracted from $\Delta N$. Then the algorithm repeats on the next nearest particle until $\delta N_p>\Delta N$. Reaching this point, the \ion{H}{II} region is sampled. The algorithm is applied to each massive star in the simulation. To help the algorithm, nearest neighbour lists of massive stars are constructed using the tree-walking routine from Phantom.

In contrast to the method described by \cite{Bisbas_2009} or \cite{dale2012}, we do not implement ray-tracing. Our method generates \ion{H}{II} regions efficiently but also produces side effects when the mass resolution is not high enough to open \ion{H}{II} regions in a very dense clump of gas or if a region starts to grow inside void (due to free boundaries in SPH). The first side effect can easily be avoided by increasing the resolution and/or attributing a fraction of the ionising temperature to the nearest particle to help the gas expand and consequently open the \ion{H}{II} region.  In the second side effect case, all ionising radiation will be focused where the gas is, as the iterative algorithm does not allow radiation escape. It could then potentially lead to an overestimate of the \ion{H}{II} region. Our implementation follows what \cite{Fujii_2021} proposed to minimise this effect by limiting \ion{H}{II} regions to a size near the initial radius of the molecular cloud. Our tests to ensure convergence of the model following \cite{Bisbas_2015} are presented in Appendix \ref{App:HII}.

\subsection{Fiducial model}\label{sec:fidu}

We studied one specific fudicial model of a giant massive molecular cloud collapse to characterise the new star formation prescription and the resulting star cluster. This model starts from an isolated isothermal massive turbulent molecular cloud. This cloud is initially spherical with a radius $R_\mathrm{c}=10$ pc and a total mass $M_\mathrm{c} = 10^4 \, \Msun$. A turbulent velocity field is initially injected into the cloud. We generated this using methods from MakeCloud \citep{grudic_makecloud_2021}. It produces initial turbulence by applying a Gaussian random field filtered by a Larson-like power spectrum \citep{larson_turbulence_1981} on the particle velocity. Compressive and solenoidal modes are introduced with $30\,\%$ and $70\,\%$ proportion, respectively. This field is finally scaled to reach a turbulent virial ratio $\alpha_t = 2$ (marginally stable condition). The cloud's initial temperature is set to $10\,\mathrm{K}$, and the mean molecular weight is set to $\mu=2.35$. This model ignores the magnetic field and external driving (pressure and turbulence). As the cloud is supposed to be isothermal, no cooling/heating processes are included, except photoionisation.

For the star formation prescription, accretion radius $r_\mathrm{acc}$ and critical density $\rho_\mathrm{crit}$ need to fulfil resolution criteria to ensure that gravitational collapse is resolved up to the sink scale. These two values then need to be chosen in agreement with the mass resolution of the simulation that gives the number of SPH particles. Our sink particle accretion radii should represent the characteristic size of a core in star-forming regions of a few thousand $\au$ \citep{sadavoy_embedded_2017}. Thus, we set the accretion radius to $4000~\au$. The critical density can then be computed using Eq.~(\ref{eq:raccrho}), which can then give a minimum mass resolution using Eq.~(\ref{eq:rhomax}) to resolve the collapse of clumps. The maximum critical density is equal to $\rho_\text{c} = 1\times10^{-18} \mathrm{g}\,\mathrm{cm}^{-3}$ using an accretion radius of $4000~\au$. The coarsest mass resolution is around $0.009~\mathrm{M}_\odot$, which gives a minimum SPH particle number of $\frac{M_\text{c}}{\Delta M} \approx 1.1$ million. Preliminary tests showed that simulations close to this minimum number gave unusual accretion behaviours. A low number of particles cannot sufficiently retrieve small turbulence modes causing faster artificial damping of turbulence, leading to early collapse. We therefore set the resolution to $3.5$ million particles for this fiducial model. Finally, we set the sink accretion time to $0.5~\Myr$, and the star seeds formation time to $0.1~\Myr$, as discussed in Sect.~\ref{sec:starform}.

Sinks are created after the density exceeds the critical density and the tests to ensure that the clump collapses pass. This means that regions of cloud exceeding the critical density are not necessarily directly transformed into sinks. These tests avoid confusing isothermal compression that can increase the gas density without gravitational collapse. However, such isothermal compression can be expensive in numerical integrations. To accelerate the computation, one can force the sink creation after exceeding the critical density by a certain factor. In this model, the sink creation forcing is set to $f_\mathrm{override} = 100\rho_c$. Table~\ref{tab:params} summarises all parameters of this model. 
\begin{table}[h]
    \centering
    \begin{tabular}{lr}
        \hline
        Parameter & Value  \\
        \hline
        $M_c$     &  $10^4 \, \Msun$ \\
        $R_c$     &  $10 \, \pc$ \\
        $\alpha_t$  &  $2.0$ \\
        $T$       &  $10 \, \mathrm{K}$ \\
        $\tau_\mathrm{ff} $ & $5.3~\Myr$\\
        $\Delta m$&  $0.00286 \, \Msun$ \\
        $r_\mathrm{acc}$&  $4000 \, \au$ \\
        $\rho_c$        &  $1\times10^{-18} \, \mathrm{g}\,\mathrm{cm}^{-3}$ \\
        $t_\mathrm{acc}$&  $0.5 \, \Myr$ \\
        $t_\mathrm{seed}$&  $0.1 \, \Myr$ \\
        $f_\mathrm{override}$& $100\rho_c$\\
        \hline
        
    \end{tabular}
    \caption{Main parameters of the fiducial model. $M_c$ is the total initial mass of the cloud and $R_c$ is its initial radius. $\alpha_t$ is the initial thermal virial ratio and $T$ is the isothermal temperature of the cloud. Its initial free-fall time is given by $\tau_\mathrm{ff} $. Each SPH particle composing the total cloud has a mass of $\Delta m$. $r_\mathrm{acc}$ is the sink accretion radius that is connected to the critical density $\rho_c$ via $\Delta m$. Respectively to the new star formation scheme, sinks can accrete during a fixed time $t_\mathrm{acc}$ and seeds inside it form after reaching $t_\mathrm{seed}$. Finally, sink creation tests can be overridden if the density reaches a too high value. This override is controlled by $f_\mathrm{override}$. }
    \label{tab:params}
\end{table}

\section{Results} \label{sec:results}

In this study, we evolved three simulations of this fiducial model with different initial turbulence states to $6.5$--$7~\Myr$. We denote the three simulations as Embedded Cluster Formation 1, 2 and 3  (hereafter ECF1, ECF2 and ECF3) in the subsequent sections. Each simulation was computed in around $14\,000$ CPU hours, which corresponds to around 18 days of computation on one compute node (32 cores).

\subsection{Qualitative description} \label{sec:quali}
\begin{figure*}
    \centering
    \begin{subfigure}[b]{0.30\textwidth}
        \centering
        \includegraphics[width=\textwidth,trim={0 0 1cm 0cm},clip]{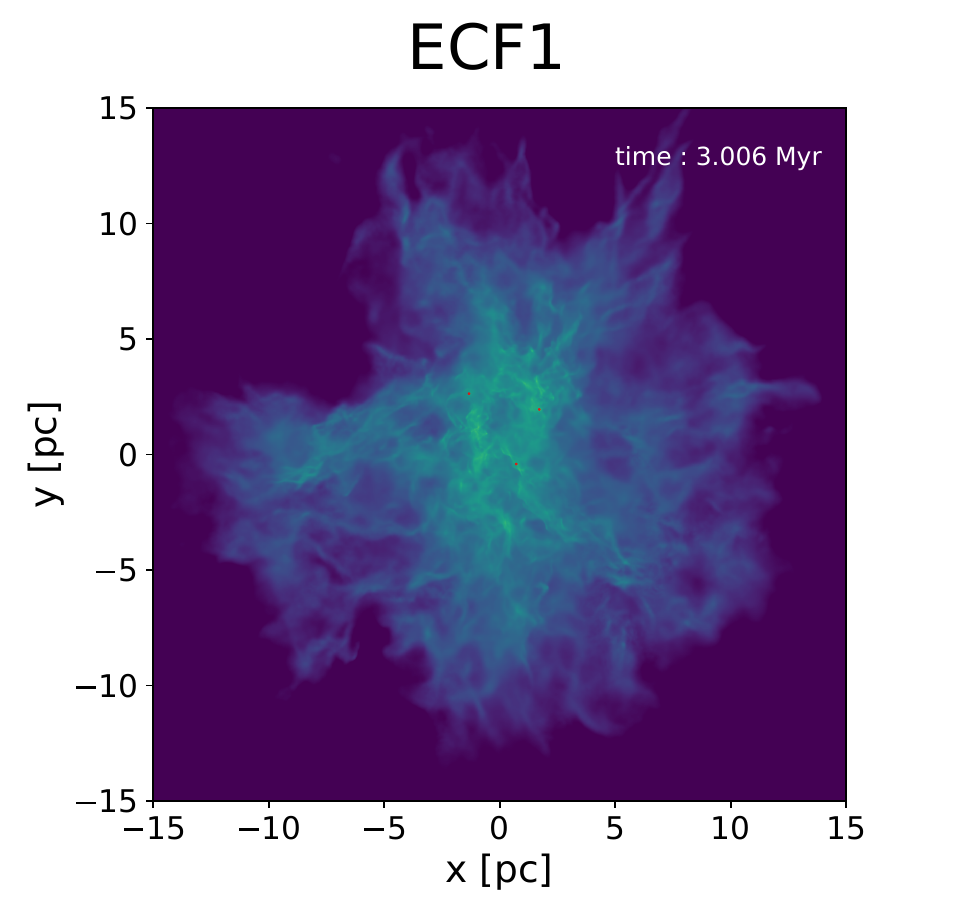}
    \end{subfigure}
    \hfill
    \begin{subfigure}[b]{0.30\textwidth}
        \centering
        \includegraphics[width=\textwidth,trim={0 0 1cm 0cm},clip]{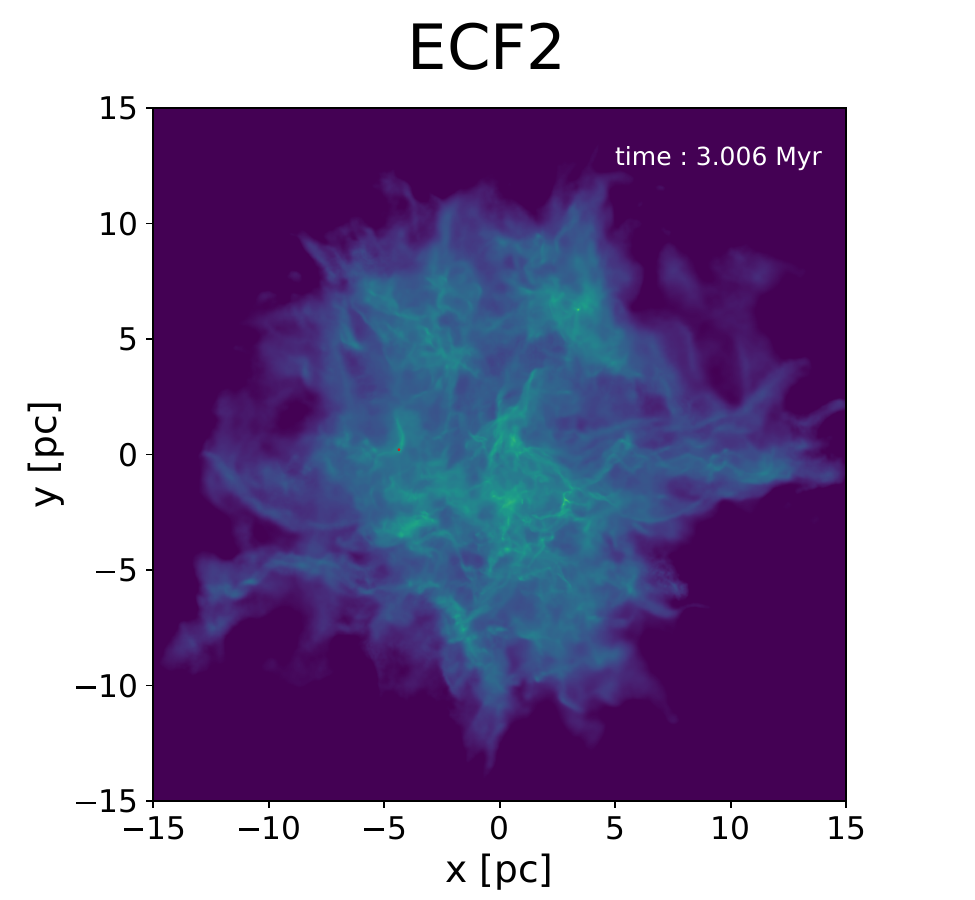}
    \end{subfigure}
    \hfill
    \begin{subfigure}[b]{0.30\textwidth}
        \centering
        \includegraphics[width=\textwidth,trim={0 0 1cm 0cm},clip]{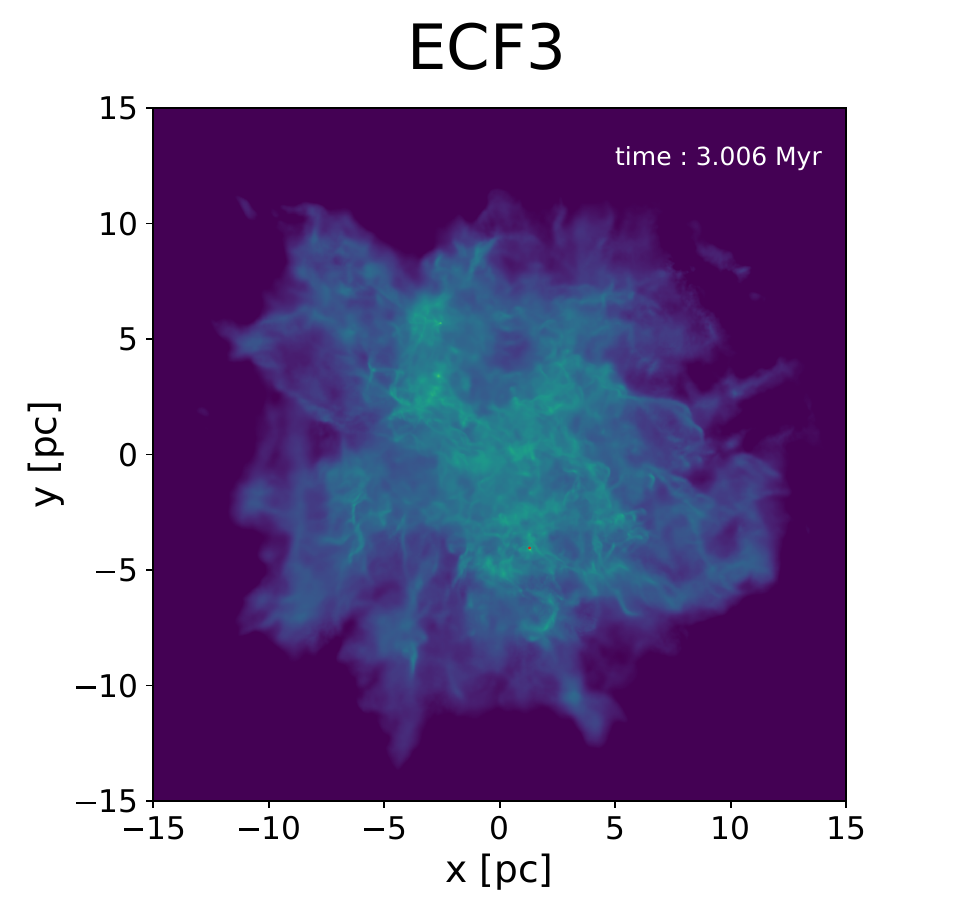}
    \end{subfigure}
    \hfill
    \begin{subfigure}[b]{0.30\textwidth}
        \centering
        \includegraphics[width=\textwidth,trim={0 0 1cm 1cm},clip]{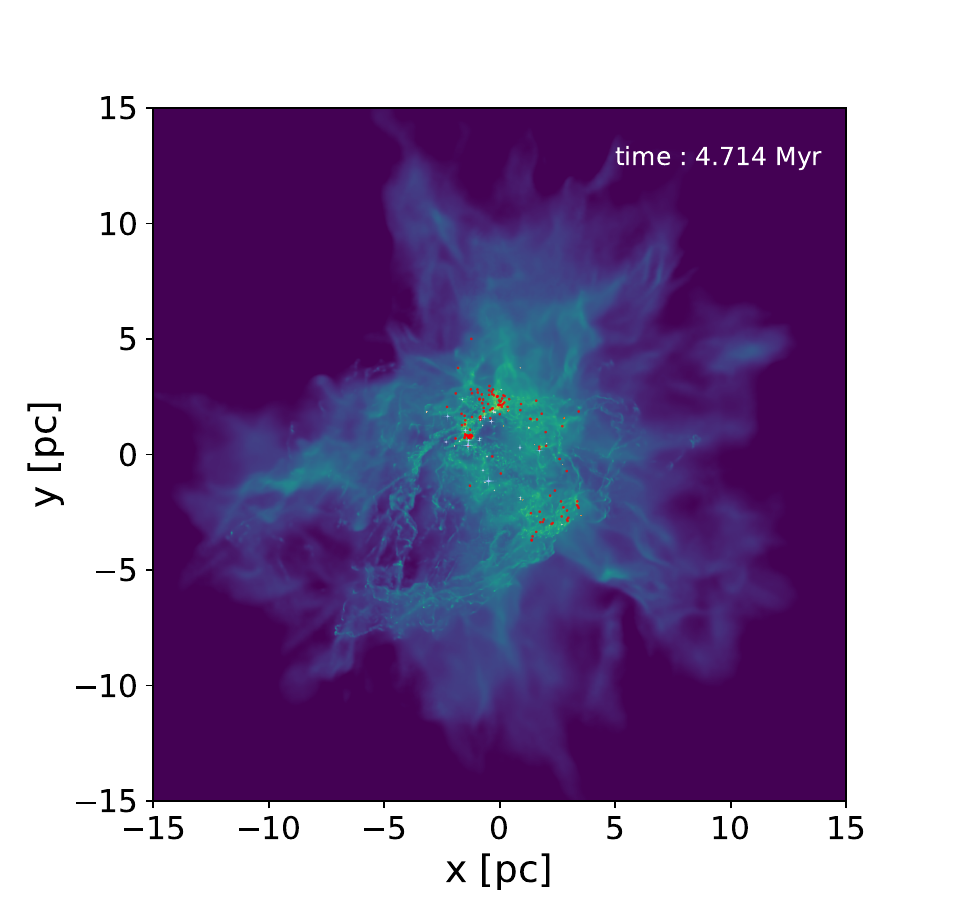}
    \end{subfigure}
    \hfill
    \begin{subfigure}[b]{0.30\textwidth}
        \centering
        \includegraphics[width=\textwidth,trim={0 0 1cm 1cm},clip]{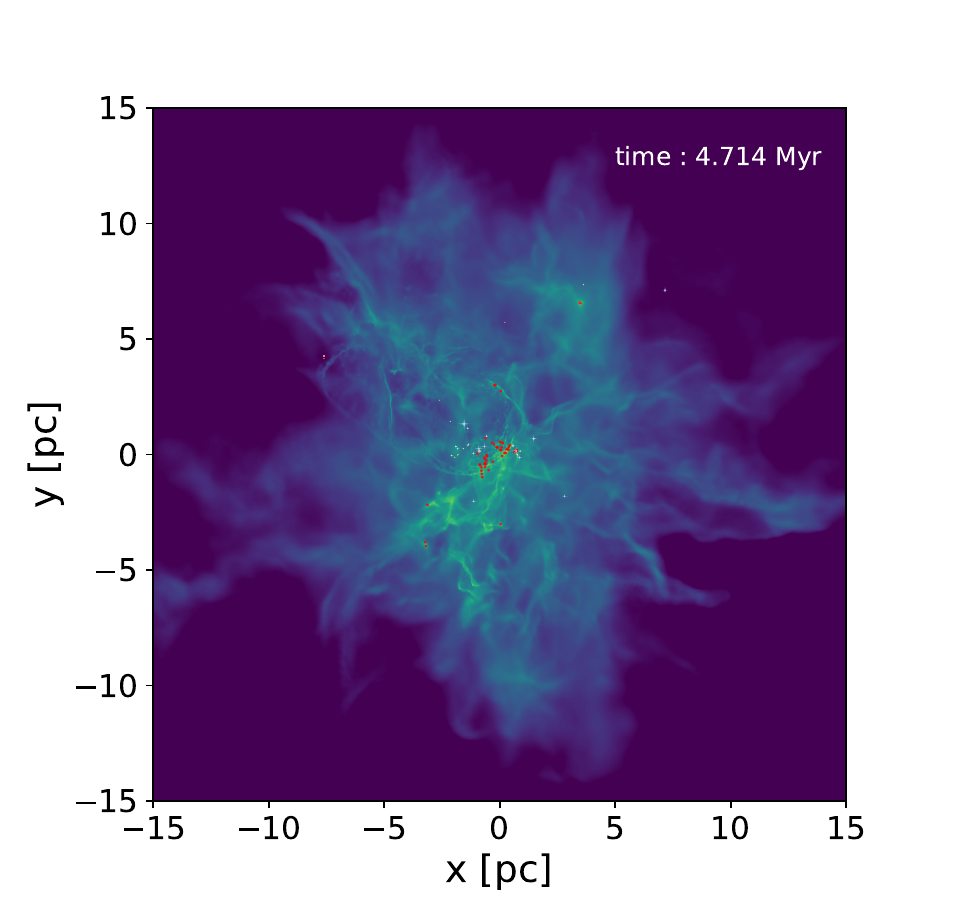}
    \end{subfigure}
    \hfill
    \begin{subfigure}[b]{0.30\textwidth}
        \centering
        \includegraphics[width=\textwidth,trim={0 0 1cm 1cm},clip]{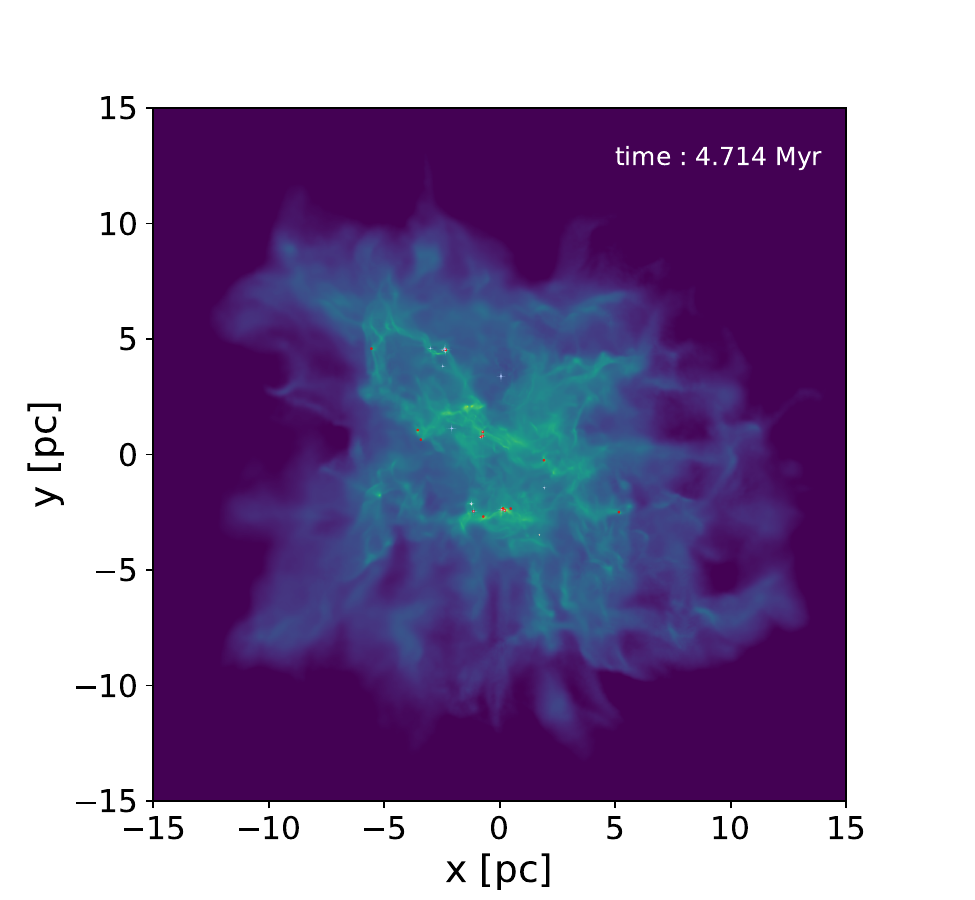}
    \end{subfigure}
    \hfill
    \begin{subfigure}[b]{0.30\textwidth}
        \centering
        \includegraphics[width=\textwidth,trim={0 0 1cm 1cm},clip]{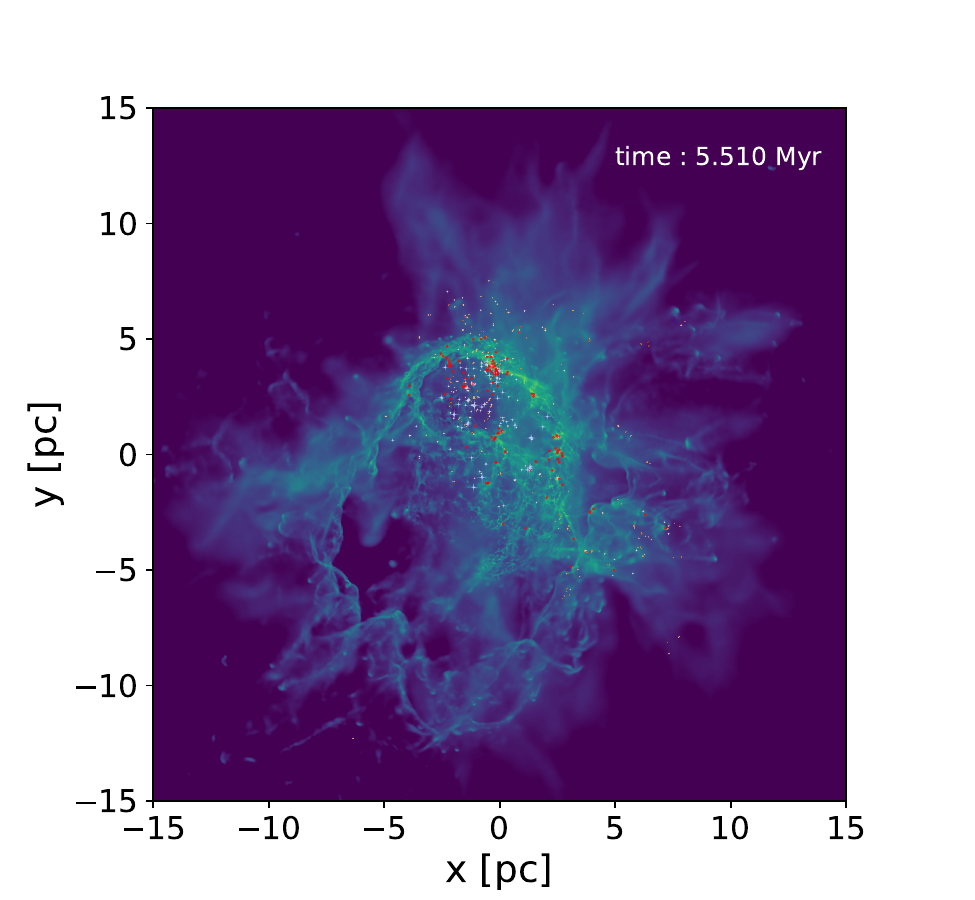}
    \end{subfigure}
    \hfill
    \begin{subfigure}[b]{0.30\textwidth}
        \centering
        \includegraphics[width=\textwidth,trim={0 0 1cm 1cm},clip]{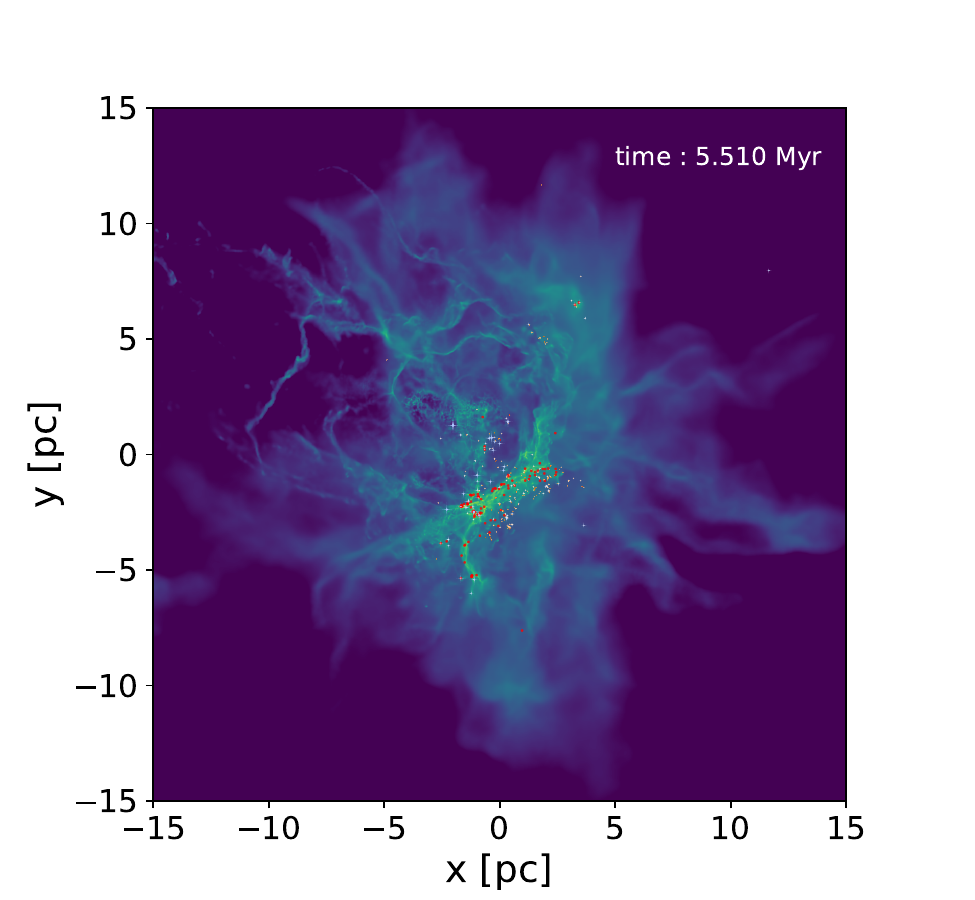}
    \end{subfigure}
    \hfill
    \begin{subfigure}[b]{0.30\textwidth}
        \centering
        \includegraphics[width=\textwidth,trim={0 0 1cm 1cm},clip]{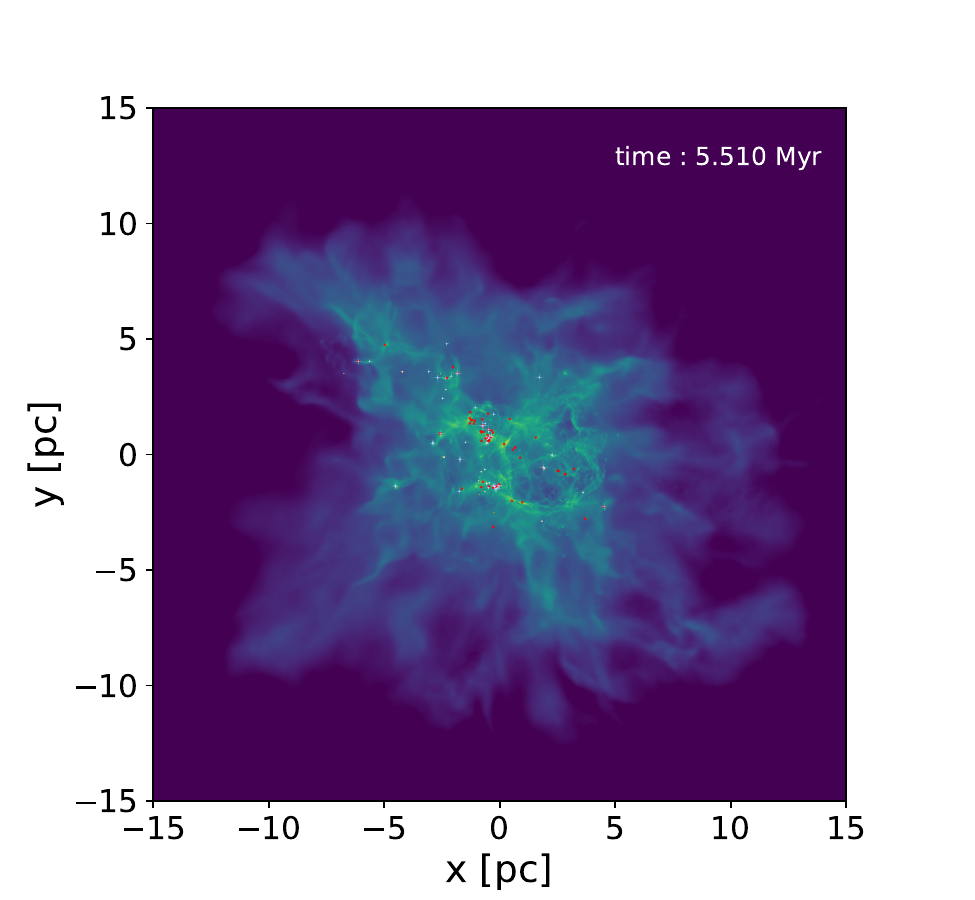}
    \end{subfigure}
    \hfill
    \begin{subfigure}[b]{0.305\textwidth}
        \centering
        \includegraphics[width=\textwidth,trim={0.88cm 0 0cm 1cm},clip]{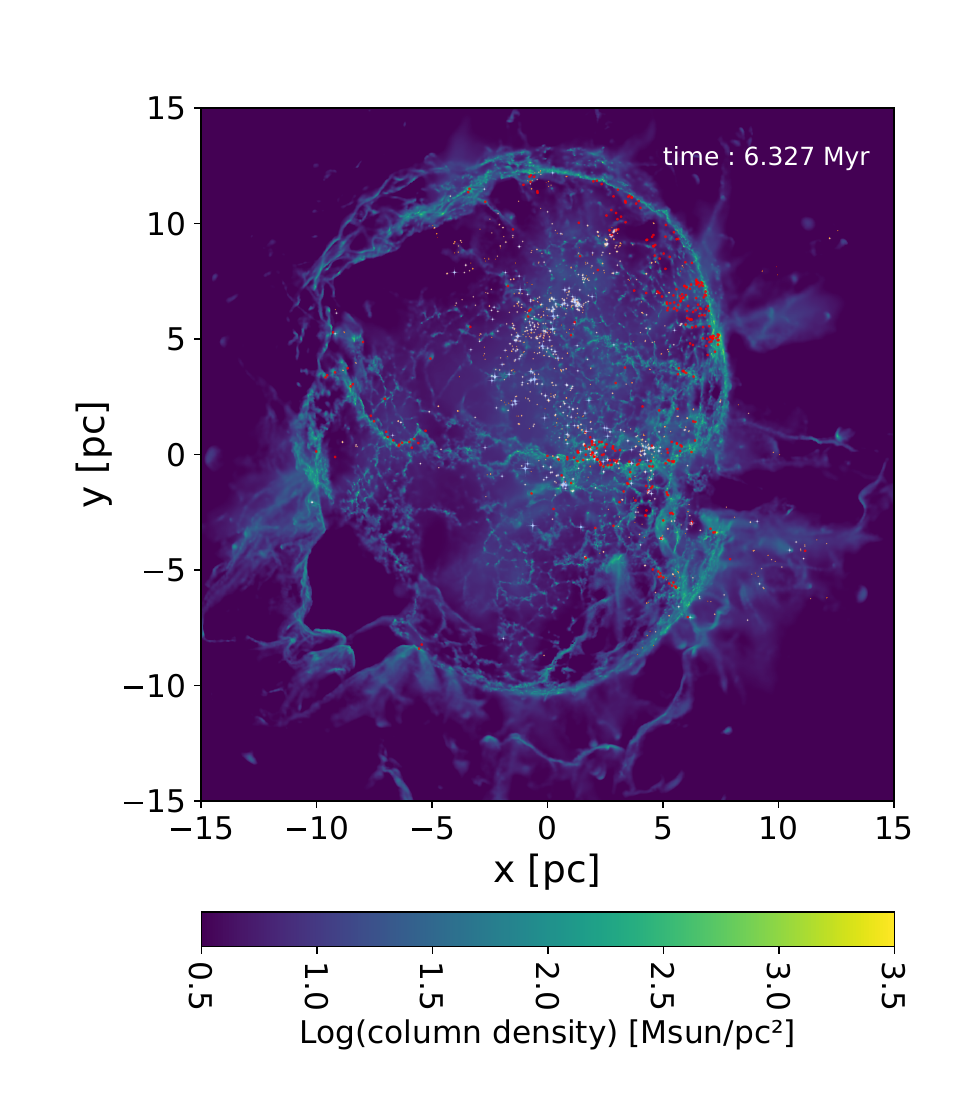}
    \end{subfigure}
    \hfill
    \begin{subfigure}[b]{0.305\textwidth}
        \centering
        \includegraphics[width=\textwidth,trim={0.88cm 0 0cm 1cm},clip]{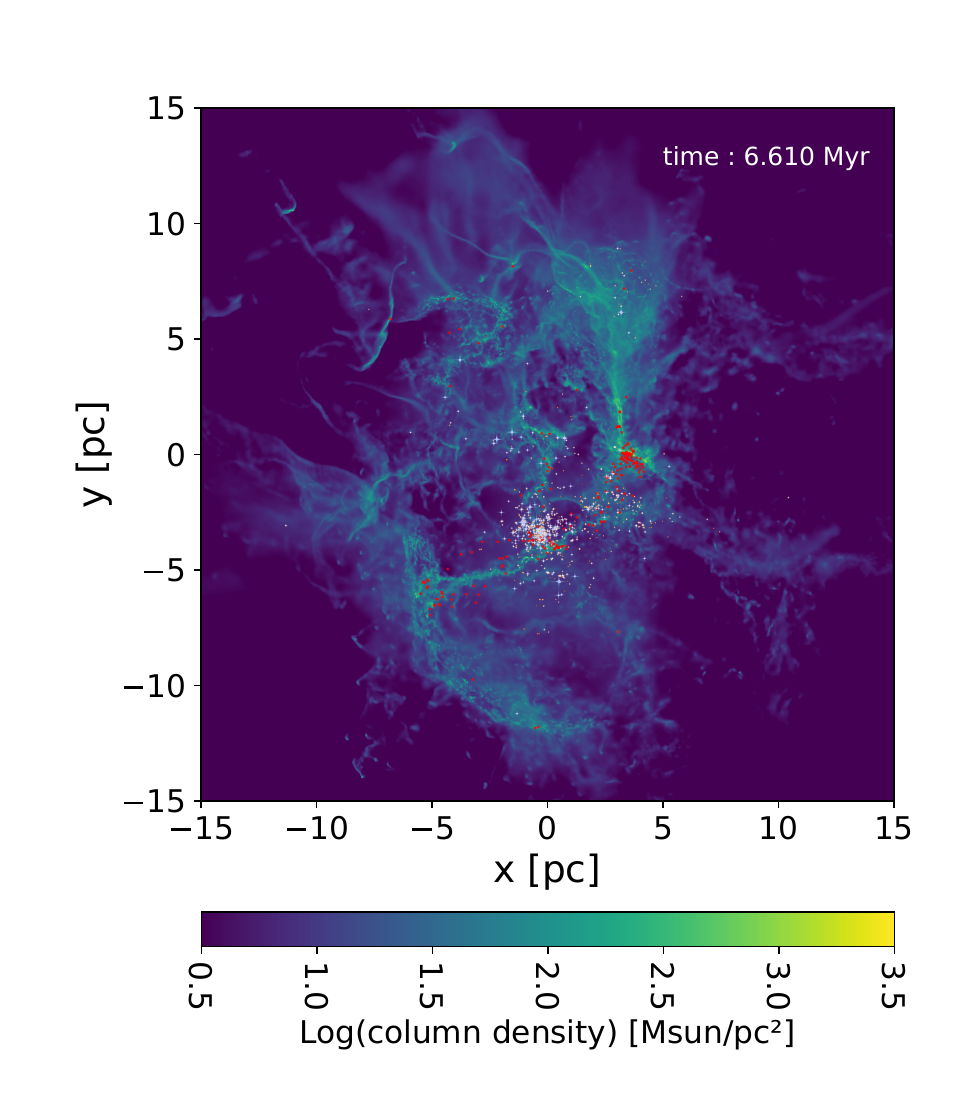}
    \end{subfigure}
    \hfill
    \begin{subfigure}[b]{0.305\textwidth}
        \centering
        \includegraphics[width=\textwidth,trim={0.88cm 0 0cm 1cm},clip]{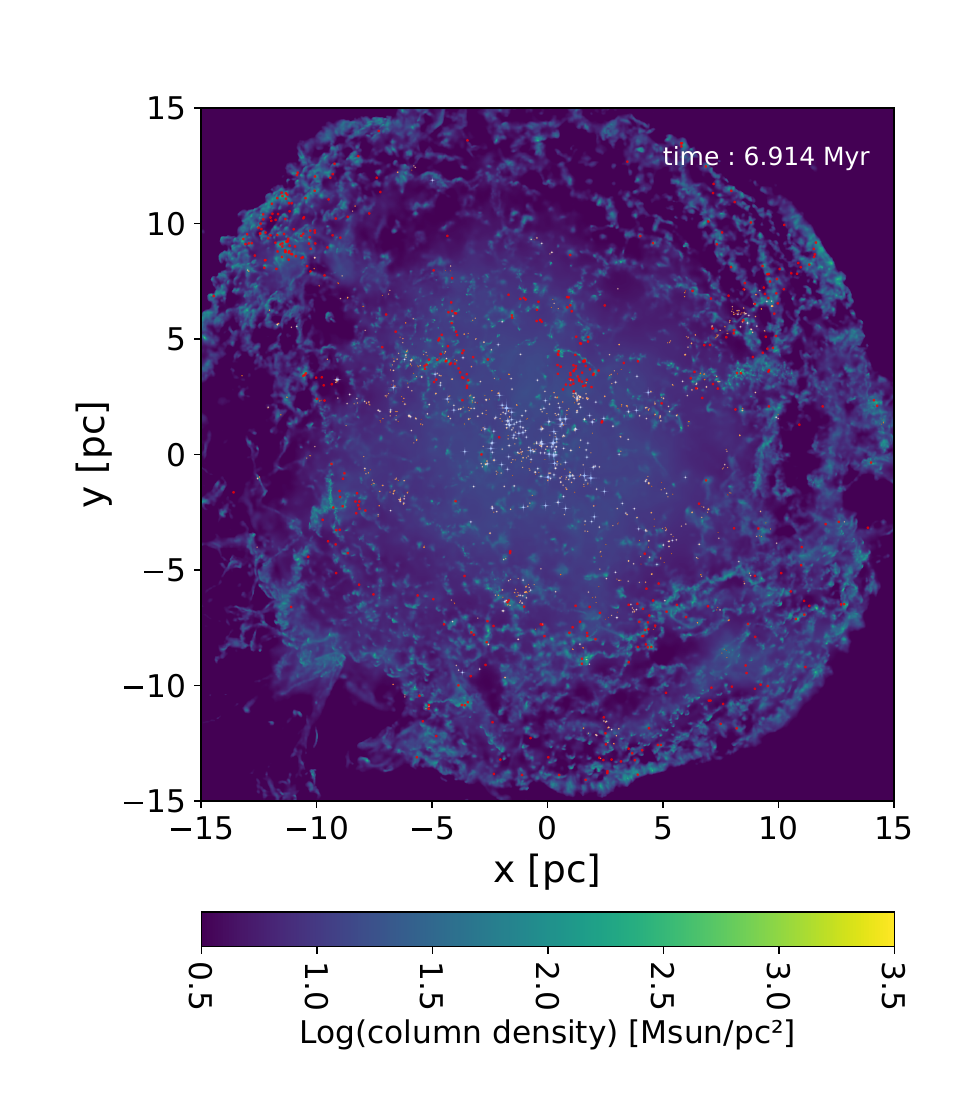}
    \end{subfigure}
    \caption{Snapshots showing the evolution of three fiducial clouds computed at multiple time steps. Each column corresponds to a specific run (top: ECF1, middle: ECF2, bottom: ECF3) and rows give different epochs in each of them. The colour map gives the $z$ integrated column density of the gas in $M_\odot\, \mathrm{pc}^{-2}$, while red circles correspond to sink positions and stars are coloured using realistic black body temperature extrapolated from a luminosity-mass relation. ECF1 presents a protostellar cluster under the influence of \ion{H}{II} regions distributed in various parts of the molecular cloud, like for instance Aquila. ECF2 presents a more centrally concentrated protostellar cluster, which develops a cluster of \ion{H}{II} regions, like the region W43. ECF3 is more difficult to link with an existing star-forming region. As all the gas has been removed very rapidly, the resulting group of stars looks like a more evolved association or cluster like NGC 6611. }
    \label{fig:ECF_snaps}
\end{figure*}

\Cref{fig:ECF_snaps} shows all three simulations at different timestamps of their evolution, from the initial gas fall up to the star cluster emergence. Each row and column corresponds respectively to a specific simulation (top to bottom) and timestamp (left to right). Snapshots display the column density of the gas overlaid by sinks and star positions.

As expected, the early evolution ($t=3~\Myr$) of each cloud is similar in each case. They start to diverge after the first star formation events. The first massive star formation presents a turning point in their evolution. These massive stars ($> 8~\Msun$) open large \ion{H}{II} regions around them. Their location and efficiency dramatically change the subsequent cloud collapse. 

The first massive star is formed early in ECF1, around $1.5 \, \Myr$ before reaching the initial free fall time. The resulting \ion{H}{II} region pushes the gas outward efficiently, inducing a significant compression in the cold molecular gas, causing it to collapse and trigger secondary star formation. In the meantime, other regions of the cloud continue their collapse and eventually produce other massive stars able to efficiently heat and ionise the cold gas with their strong radiation. Their accumulation accelerates the gas expulsion, leaving behind a new cluster of stars exposed from their parental cloud. Triggered formation following expanding bubbles is imprinted with their motion, resulting in a global expansion of the newly formed cluster.

ECF3 shows similar outcomes from ECF1, with a large expanding \ion{H}{II} region hosting secondary formation at its outskirts. However, gas expulsion is more efficient. The final cluster appears correspondingly looser and more spatially extended. It did not produce massive stars early. Instead, the cloud collapse had more time to concentrate its mass in a central region. This stronger collapse produces multiple massive stars in a short time interval and in the same area. The coherent addition of \ion{H}{II} regions leads to the most efficient gas expel and the highest cluster expansion rate seen in this work.

ECF2 produces a highly off-centred intermediate massive star early in the simulation in comparison to the other two. Its \ion{H}{II} region expands slowly and pushes the gas gently, helping it to collapse naturally under its weight. Moreover, later-formed massive stars do not show coherent effects on their \ion{H}{II} regions as in ECF3, even if these stars are the most massive ones in all simulations. Instead, shocks are produced between \ion{H}{II} regions, thereby facilitating the condensation of cold gas to an even greater extent. With the help of gravitational collapse, this leads to the formation of a very compact cluster that is not yet fully revealed from its natal gas and where star formation is still ongoing at the end of the computation ($\sim6.6 \, \Myr$). 

 The triggered formation has already been seen and studied in numerical simulations. \cite{dale2012} performed molecular cloud collapse with \ion{H}{II} feedback relatively similar to ours. They found in their simulation that ionising radiation from massive stars shapes the gas inside such molecular clouds by creating large bubbles and pillars, which are also observed here. They also concluded that this radiation feedback induced a significant rise in star formation, which also corroborates what is found in our simulations. However, their calculation stops rapidly (Model J) due to growing time step constraints on the stellar dynamics. Our new methods allow us to continue the calculation further. Thanks to this amelioration, we can now see that this triggered star formation can still happen far from the revealed cluster, creating an expanding distribution of low-mass stars around denser clusters.
 
 The formation of massive stars, their spatial locations, and their coherent contributions are key elements in the future of a young stellar cluster that will be sculpted by \ion{H}{II} region expansion. Their locations and masses are function of the initial turbulent velocity field and mass sharing inside sinks, which is random by definition. It means that the formation of such clusters is, like their evolution, highly stochastic. This confirms that a statistical study is necessary to identify the main evolutionary paths of embedded clusters.

\subsection{Star formation efficiency} \label{Sec:SFE}
The gas mass conversion and its evolution in time can help characterise the star formation inside clouds. Dividing the converted mass by the initial gas mass gives an efficiency factor referred to as the \emph{star formation efficiency} (SFE). This coefficient is an indicator of the evolutionary stage of a star-forming region and is computed at various scales in observational works, from 0.01 to 10~pc \citep{Alves_2007, Evans2014, Louvet_2014, Konyves_2015}. From an entire giant molecular cloud to a smaller scale, like in the neighbourhood of dense cores inside a larger complex, the definition of the SFE is the same and is given by
\begin{equation}
    \mathrm{SFE} = \frac{M_\star}{M_\text{cloud}+M_\star},
\end{equation}
where $M_*$ and $M_\text{cloud}$ are the total mass of stars and the total mass of gas inside the considered region, respectively. In our study, star formation efficiency is computed at the cloud scale. Hence, no external mass is added to the calculation, which may not be the case in observations due to external gas inflows and protostellar outflows.

\begin{figure*}
    \centering
    \includegraphics[width=0.9\paperwidth]{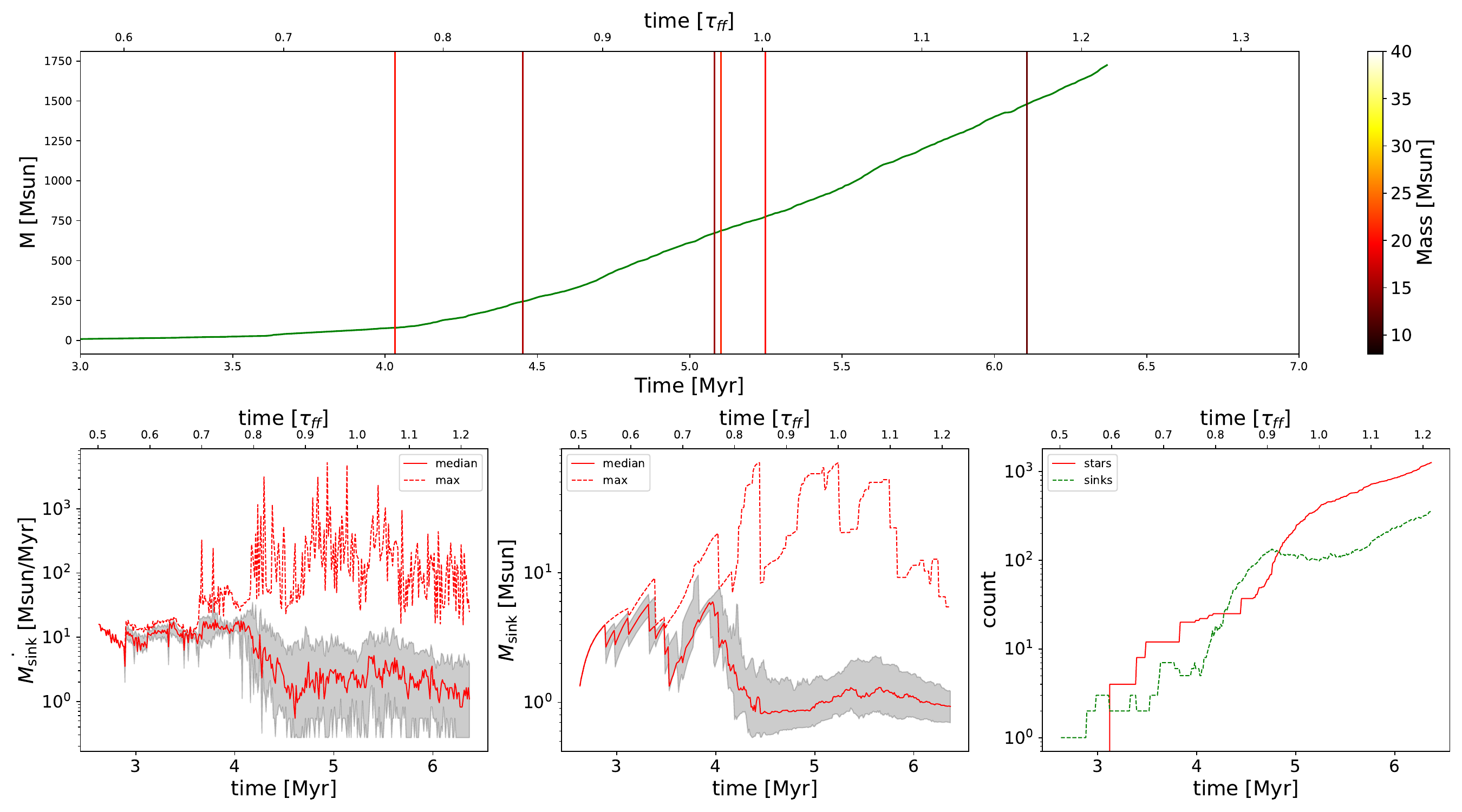}
    \caption{Evolution over time of multiple key parameters linked to the gas-to-star mass conversion. The top panel shows the time evolution of the mass of gas converted in sinks and stars in ECF1. The formation of massive stars is represented by vertical lines, with their colour indicating the mass of the massive object. The bottom panels show the temporal evolution of the accretion rate (left), mass median (centre) and the number of stars and sinks formed in the simulation (right). For the first two panels, the quartile range (grey zone) and the maximum value (dashed line) are also shown. It allows us to see that the sharp transition at $4 \Myr$ occurs on the majority of sinks. However, even after the change of regime, the cloud is still capable of producing some massive sinks with a high accretion rate.}
    \label{fig:SFE1}
\end{figure*}

\begin{figure*}
    \centering
    \includegraphics[width=0.9\paperwidth]{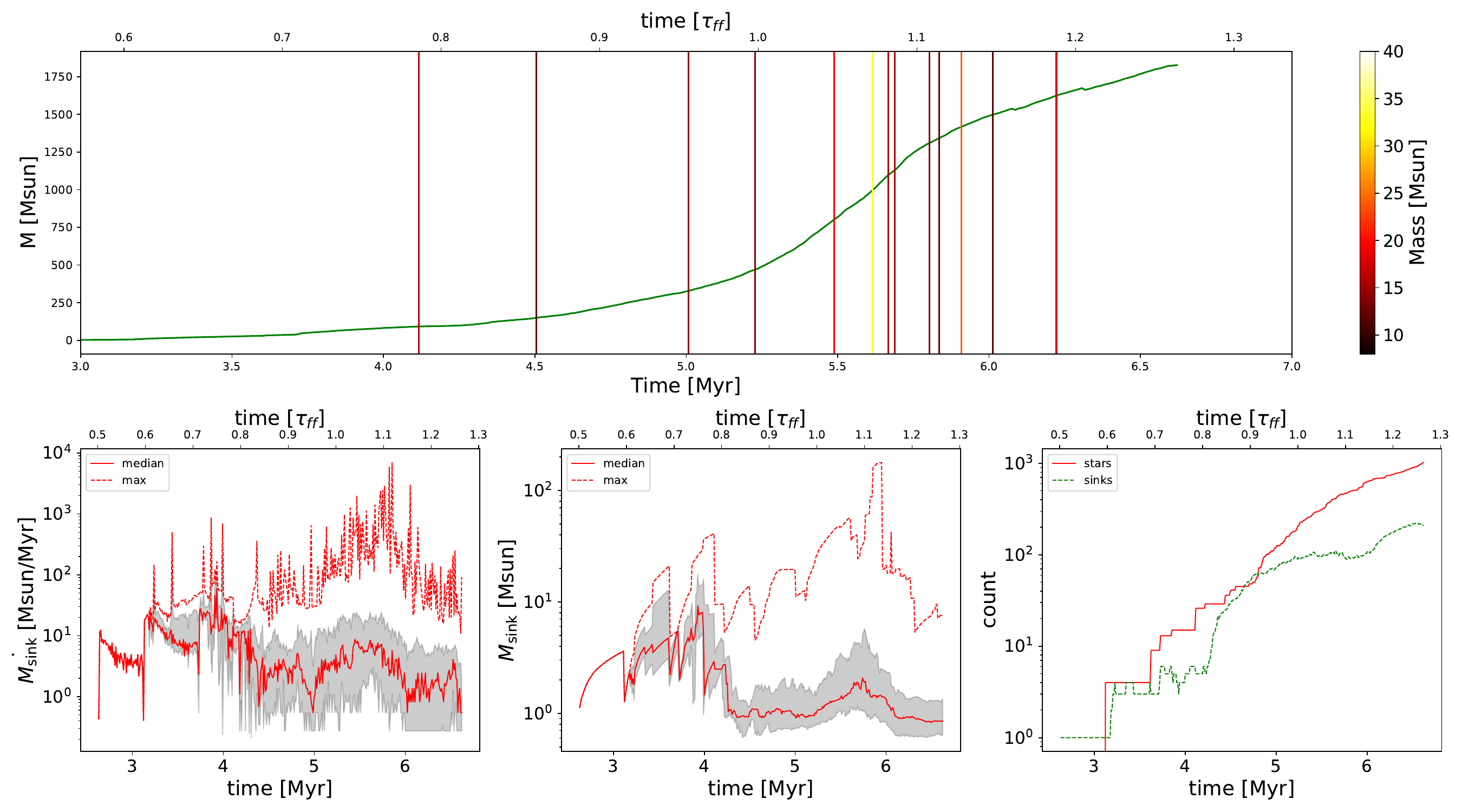}
    \caption{Same as previous figure applied on ECF2 simulation}
    \label{fig:SFE2}
\end{figure*}

\begin{figure*}
    \centering
    \includegraphics[width=0.9\paperwidth]{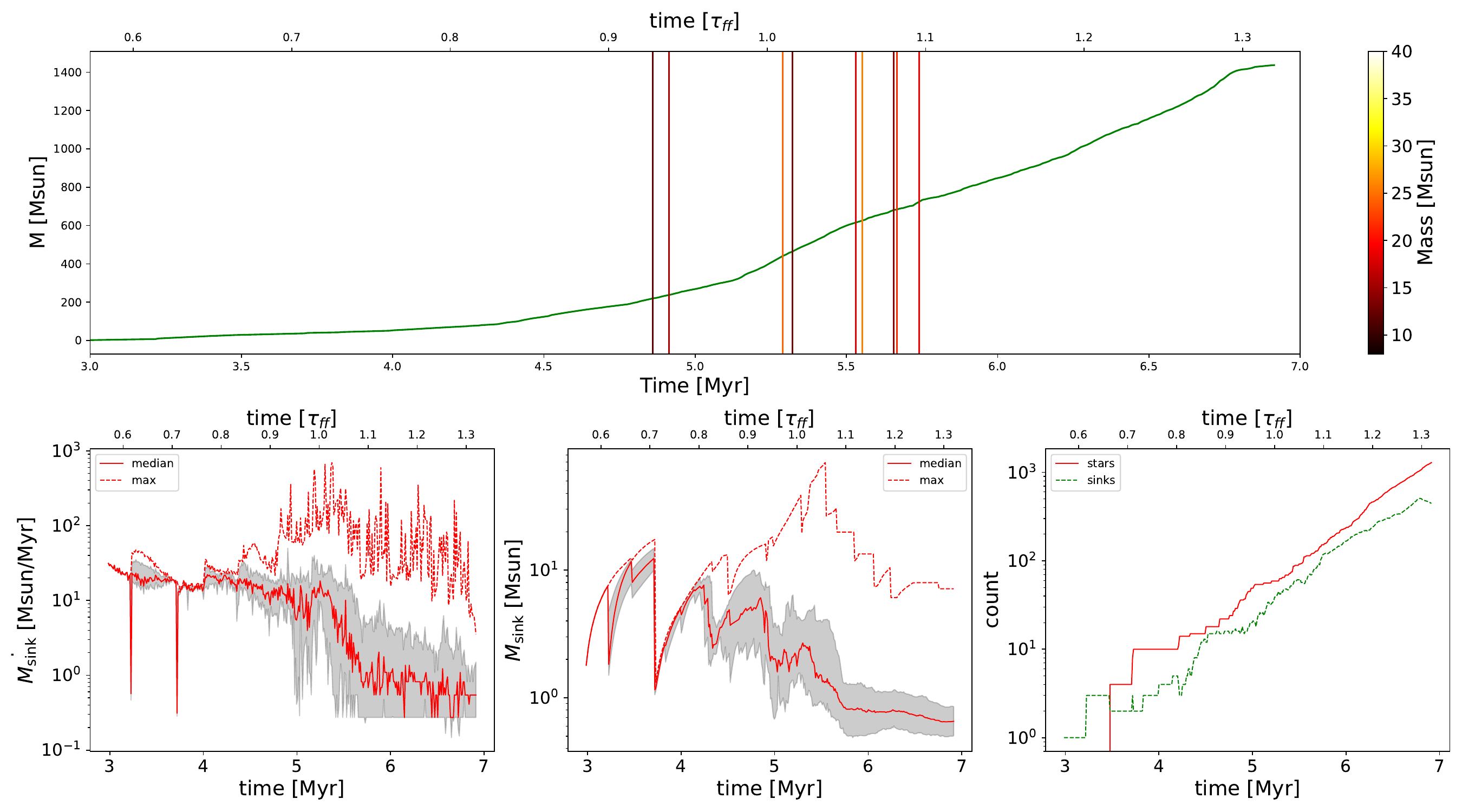}
    \caption{Same as previous figure applied on ECF3 simulation}
    \label{fig:SFE3}
\end{figure*}

Figures~\ref{fig:SFE1}, \ref{fig:SFE2}, and \ref{fig:SFE3} show for each simulation the evolution over time of key parameters linked to the gas mass conversion into stars, namely the star formation efficiency (top panel), the median values of sinks mass and accretion rate (bottom left and centre), and the number of sinks and stars (bottom right). Each simulation reaches a SFE exceeding $10\%$, with values between $14 \%$ and $ 18 \%$. These efficiencies are high compared to those observed in parsec-scale star-forming regions \citep[see however][]{Louvet_2014}. Such high efficiencies tend to be found at much smaller scales \citep[e.g.,][at 0.01~pc scales]{Konyves_2015}. This suggests that other feedback mechanisms, such as jets \citep{Machida_2012}, or other physics such as non-ionising radiation and magnetic fields \citep{price_2009}, may be needed to regulate the mass conversion. However, it is difficult to perform a direct comparison as real molecular clouds are certainly not in complete isolation and can gather gas inflow, which will reduce the SFE.

Most of the gas conversion occurs in $\sim2 \, \Myr$, between $4$ and $6 \, \Myr$ ($0.75$ to $1.13 \,\tau_\text{ff}$). The first sinks appear around $3 \, \Myr$ ($0.56 \, \tau_\text{ff}$) and are limited to a small number during one million years. The mean star formation rate (SFR), given by the mean slope of the mass conversion evolution in time, is equal to $6.8 \times 10^{-4} \, \Msun \, \yr^{-1}$ for the first two runs. ECF3 shows a slightly lower value with $4.5 \times 10^{-4} \, \Msun \, \yr^{-1}$. All three mass conversion slopes increase monotonously during the whole computation.

It is difficult to see where star formation halts from the number of sinks over time. In ECF2 and ECF3, the number of sinks starts to decrease at the last few points. This, in addition to the SFE slowdown in both simulations, suggests that star formation is about to stop. By contrast, the number of sinks is still increasing in ECF1 and the star formation efficiency does not show any slowdown. This late increase is only due to secondary star formation in \ion{H}{II} regions outskirts. It could also be induced by numerical effects. It seems clear that simulations should be evolved further in time to determine when star formation stops. (discussed in \cref{sec:discussion})

A key event occurs between $4$ and $4.5 \, \Myr$ ($0.75$ to $0.84 \,\tau_\text{ff}$) for all calculations, when the star formation process produces a sharp rise in the number of sinks produced in the cloud. Except for ECF3, this moment coincides with the first formation of massive stars.

The median values of sink mass and accretion rate show two different regimes at the same turning point of $4 \Myr$ for ECF1 and ECF2. In ECF1, before this time, the mean value of the median for both plots is equal to $10~\Msun \, \yr^{-1}$ and $3~\Msun$. These values drop to $2~\Msun \, \yr^{-1}$ and $0.8~\Msun$ just after the turning point. The number of sinks also drastically increases from less than ten to more than a hundred in less than a million years after this timestamp. This sudden jump occurs due to the opening of an \ion{H}{II} region. 

Expanding bubbles in these simulations greatly modify the star formation regime making a sharp transition from a dominant massive formation channel to a low-mass one, as observed in some evolved massive protoclusters \citep[e.g.][]{armante2024}. This new channel produced stars in the outskirts of \ion{H}{II} bubbles where sinks and gas rapidly decouple after their formation. As sinks form, these only evolve through gravitational interaction and lose pressure support that drives \ion{H}{II} bubble expansion. Thus, if this expansion accelerates, sinks do not have the time to accrete material before entering an ionised region where accretion is no longer possible. Most of the converted mass after the turning point then goes mostly into low-mass sinks, explaining the sharp transition.

In unperturbed locations of the cloud, massive stars still form following the natural cloud gravitational collapse. A revival of both accretion rate and mass can be seen to start at $\sim0.95\,\tau_\text{ff}$ in ECF1 and ECF2. And it reaches a maximum at $\sim1.05\,\tau_\text{ff}$ that coincides with the most massive sink formation in both simulations. Higher revival in ECF2 also agrees with previous observations (see \cref{sec:quali}) which showed that \ion{H}{II} bubble expansion was more gentle, leading to a forced collapse instead of a dismantling. ECF2 produces the most massive sink and then the most massive star with a mass of $180~\Msun$ and $45~\Msun$ respectively. It seems logical that regions unaffected by \ion{H}{II} bubbles will collapse very rapidly and efficiently near the initial free-fall time. The behaviour of the ECF1 and ECF2 runs reminds the Aquila and W43 protoclusters \citep{konyves2015, motte2003}.

In comparison, ECF3 did not produce early massive stars like the other two. This particular simulation shows a smoother formation process up to $\tau_\text{ff}$, maybe resembling the star formation activity in nearby, low-mass star-forming regions. It follows the gravitational collapse of the cloud that naturally accelerates. Thus, the number of formed sinks increases smoothly. The lack of forcing by the \ion{H}{II} region prevents the sharp transition at $0.75 \,\tau_\text{ff}$ as in the other runs. The acceleration of the collapse stops sharply at $\sim1.05\,\tau_\text{ff}$. As many massive sinks were able to cluster in dense regions without any feedback, this led to the simultaneous formation of massive stars in the same region. This burst of massive star formation produces the sharpest regime transition observed in the simulations, permitting solely low-mass star formation until the end of the computation. However, this scenario is probably not the sole possibility if early massive star formation does not occur. Finally, the star formation efficiency of this simulation is the lowest, which may be due to rapid gas removal. This may correlate with a conclusion made by \cite{dale2012}, who link lower star formation efficiency to efficient photoionisation feedback.

\section{Discussion} \label{sec:discussion}
\subsection{Numerical performance}

In terms of computational efficiency, we found our new N-body and \ion{H}{II} region implementations to be effective at speeding up the calculation when the stellar environment densifies. Our new subgroup methods correctly handle every close encounter or binary. ECF2, which shows the densest cluster formed in this study, produces a maximum of 17 subgroups involving 38 stars at the same snapshot. Even if this number is small compared to the total number of stars at the end of the simulation ($N_{final}=1225$), the performance boost is significant, as only one of these subgroups can impose a timestep that is orders of magnitude lower in the substep loop. We indeed observe one bound group which has a time step a hundred times smaller than the minimal timestep with no subgroup integration with a minimum separation of $4 \, \au$.

To properly demonstrate this performance boost, we relaunched the simulation near the time of creation of this hard binary. This new calculation took 48 hours (walltime) to complete only one-fifth of the time integration until the next dump. In comparison, the whole time integration took $3.1$ hours with our new implementation, which gives a gain of approximately 80. The gain depends on the hardest binary in the system.

With our methods, time step constraints set by individual stars (not in subgroups) are always close to the CFL (Courant-Friedrich-Lewy) constraint \citep{courant1928} of the densest SPH particles in the cloud during the simulation, which sets the slow force minimum time step, i.e., the number of substeps is always contained to a reasonable number (a few tens at maximum). Hence, none of these parts of the calculation is a major bottleneck.

The main performance hit is due to the interactions between point masses (sink and star) and the gas. These are calculated with a direct method in Phantom and always at the lowest time step of the scheme. In most applications of Phantom to date, this part of the algorithm was not the main bottleneck, as the sink number was not large enough for the $\mathcal{O}(N_\mathrm{sink}N_\mathrm{gas})$ interaction to dominate the complexity. Such interactions essentially scale linearly with particle numbers in our simulations, like in previous studies. As a result, this bottleneck makes the computation too expensive to go further in time than 6 to 7 Myr for our application. 

Since the number of point masses is close to three thousand in our simulations, these interactions start to dominate the computational expense. As said above, they are also computed for every point mass at every time step, even when only a few gas particles are active in the individual time stepping scheme. It is a waste of computation time, as the time step constraints of sink/star and the gas interactions are often two or three orders of magnitude higher than the smallest one. On average, this part of the calculation represented more than $80\,\%$ of the computation time of one global step when the number of sinks is higher than $\sim 1000$. This bottleneck is the main remaining limitation that makes it difficult to carry out this type of simulation on a statistical scale, which is the ultimate goal of the project. Future work will focus on optimising this part of the calculations. For example, following what has already been done in collisional N-body codes \citep{Mukherjee2021,wang_petar_2020}, these forces --- as they do not strongly perturb the evolution of hard groups or close encounters --- can be computed with the help of the Phantom kd-tree reducing the complexity to $\mathcal{O}(N_\mathrm{sink}\mathrm{log}(N_\mathrm{gas}) + N_\mathrm{gas}\mathrm{log}(N_\mathrm{sink}))$ and/or by only computing these forces at the first integration level. The latter solution would make it possible to avoid computing these interactions multiple times during the substep scheme, and computing them only once per step should save most of the unnecessary calculations.

\subsection{Effect of $f_\mathrm{override}$ on sink formation output} \label{sec:fovr}

We investigate the effect of the $f_\mathrm{override}$ parameter used to force the creation of a sink if an SPH particle reaches the critical density without passing other physical tests detailed in \cite{price_phantom_2018} or \cite{bate_modelling_1995}. These tests check whether the gas clump is indeed starting to collapse gravitationally, or whether it is just experiencing hydrodynamical compression. If the gas is isothermal, such compression can rise several orders of magnitude higher than the critical density. In our simulations, high density implies high time constraints which will drastically slow down the computation. Moreover, all this extra cost is unnecessary, as the critical density is supposed to be close to the maximum resolution of the simulation.  The extra computation above the critical density is not guaranteed to be physically accurate. $f_\mathrm{override}$ is therefore here to force the creation of a sink after reaching 
\begin{equation}
    \rho = f_\mathrm{override} \times \rho_\mathrm{crit},
\end{equation}
with $f_\mathrm{override}$ equal to several orders of magnitude to find a good compromise between efficiency and accuracy.

For the sake of efficiency, this parameter was set to 100 in the simulations of this paper, but this is a relatively low value. With this value, HII region expansions produced in the simulations showed efficient sink formation at the outskirts of these regions. The compression induced by hot bubbles in the cold collapsing gas can force this collapse and lead to a burst of sink production. However, it can also produce compression in non-collapsing regions that can exceed the critical density. Therefore, we find that a too low $f_\mathrm{override}$ parameter could overestimate this burst of sink production, leading to an overestimated number of low-mass stars at the end.

\begin{figure*}
     \centering
     \begin{subfigure}[b]{0.49\textwidth}
         \centering
         \includegraphics[width=\textwidth]{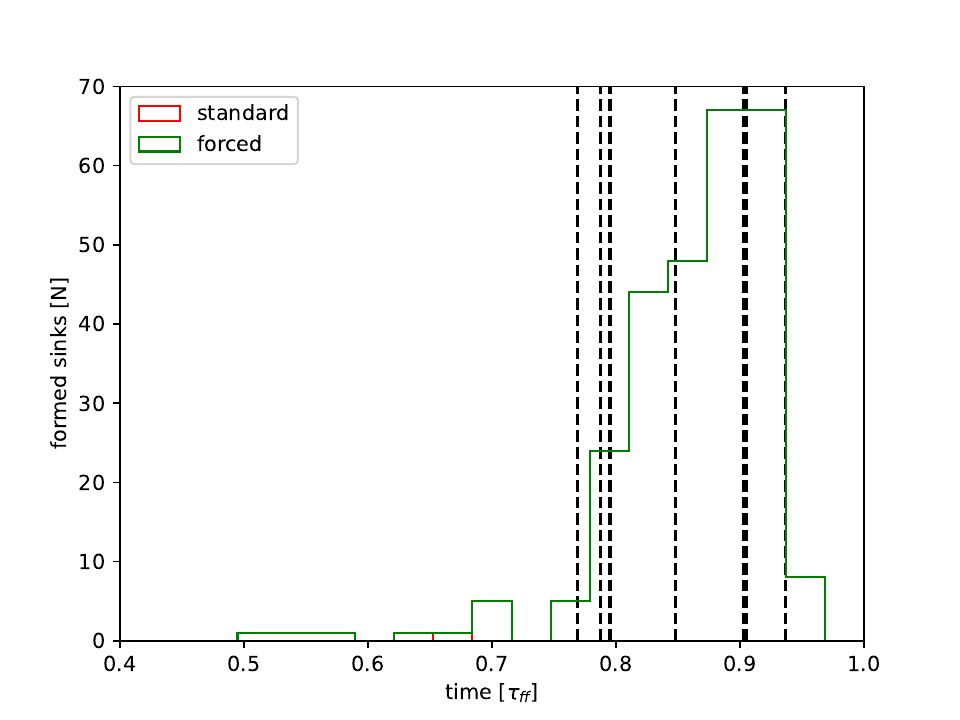}
     \end{subfigure}
     \begin{subfigure}[b]{0.49\textwidth}
         \centering
         \includegraphics[width=\textwidth]{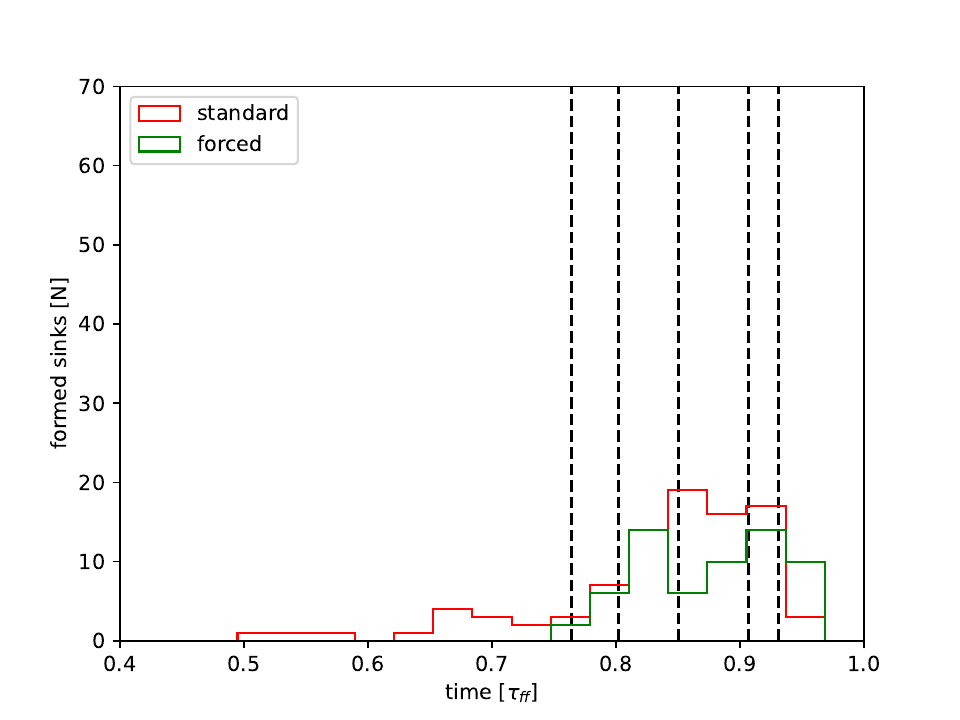}
     \end{subfigure}
        \caption{Histograms of the number of sinks over time. The red histogram gives the formation of sinks passing every test. The green histogram gives the forced sinks. Vertical dashed lines represent massive star formation. ECF1 is on the left panel and the right panel is the same initial conditions with $f_\mathrm{override}$ set to 1000 referred to as ECF1B.}
        \label{fig:fovr}
\end{figure*}

Figure~\ref{fig:fovr} shows histograms of the formation of sinks over time, and up to one freefall time, of two simulations with different $f_\mathrm{override}$ values (ECF1 with $f_\mathrm{override}=100$ and ECF1 with $f_\mathrm{override}=1000$). ECF1 produced all its sinks, except one, via forced creation. Looking at ECF1B, we see that the first sink creations are similar. Thus, the forced creation does not overestimate the production of sinks during the initial collapse. The sink productions of ECF1 and ECF1B start to diverge after the first massive star formation. The total number of sinks produced after this timestamp is 263 (only forced) and 141 (79 standard and 62 forced) for ECF1 and ECF1B, respectively. Compared to ECF1B, ECF1 overestimates by at least 1.9 times its sink production at the outskirts of \ion{H}{II} regions with $f_\mathrm{override}$ set to 100. While some of the forced sink creations are physical, and not numerical artefacts, a higher value of $f_\mathrm{override}$ helps to minimise this numerical side-effect. Therefore, $f_\mathrm{override}$ should be set higher than 100 in future simulations while keeping numerical efficiency in mind.

\subsection{Sink merging}

As mentioned in Sect.~\ref{sec:starform}, sinks can collide with others during their accretion phase. If two sinks collide and both encompass their centre of mass, they fulfil the conditions for a sink merger. In such a case, the two sinks are merged to form a new sink that inherits the physical properties of its parents. The new position and velocity are the barycentres between the two parents. This ensures the conservation of linear momentum during the simulation. The angular momentum is also conserved as Phantom tracks the spin of all sinks in the simulation. The sink properties provided by our new star formation prescription are also updated during a merger (see \cref{sec:starform}). In particular, the number of seeds in the initial sinks is summed into the merged sink.

Mergers are quite common in these simulations. The gravitational pull of the cloud tends to bring sinks closer over time, which inevitably creates collision trajectories between them. Mergers are key to forming very massive sinks that can then create massive stars. This merging process can take place consecutively on the same sink, leading to very massive sinks (the most massive one reaches $180 \, \Msun$ in ECF2). Accretion rates measured in these simulations are too low without merger events to explain the formation of stars of $20$ to $40~\Msun$, which are essential to rapidly remove the gas from the star-forming region with their \ion{H}{II} feedback. 

In these cases, summing the parents' star seeds can bring their number to a few dozen in a single sink (50 at maximum in ECF2). This can be seen as a contradiction with our main prescription hypothesis that permits only a few star seeds (less than five in these simulations) in a sink. This means that the most massive sinks need to share their mass with a large number of stars during the release phase. Randomness during sharing is smoothed in such cases as it needs to cut the parent mass in more smaller parts. As a result, most massive sinks create a large number of intermediate-mass stars. This is at odds with our model, which predicts that the most massive stars are produced inside very massive sinks. Therefore, our random sharing of mass between star seeds can bias the final star mass spectrum by undersampling high-mass stars and oversampling intermediate-mass stars. Since less massive stars also imply less feedback, it can also have an impact on the star formation efficiency and dynamics inside the star-forming region. Ultimately, we need to improve the merging process in relation to our new star formation scheme to make sure that the number of seeds inside a sink is always in agreement with the maximum value allowed, e.g. by adding star escapers after a merging and/or modelling seeds merging inside our subgrid model.

\section{Conclusions}\label{sec:conclusion}
This paper presents a new implementation of direct $N$-body algorithms as well as prescriptions for star formation and stellar feedback in the SPH code Phantom. Together, they unlock the possibility of studying embedded star cluster evolution from their birth inside a molecular cloud to their unveiling from their natal gaseous environment. The new algorithms come in two parts:

First, our new star formation prescription was implemented, which provides distributions of stars self-consistently formed from molecular cloud collapse at low computational cost. It uses thousand-$au$-diameter sink particles that host star seeds within them and are transformed into stars after passing some criteria.

Second, we completely refactored the point mass time integration methods in Phantom. We replaced the existing second-order Leapfrog integrator with an accurate fourth-order Forward Symplectic Integrator (FSI), which is now the default in the public code. Its symplectic properties guarantee perfect conservation of angular momentum and good conservation of energy. Its forward properties also guarantee that there is no issue with dissipative forces in the code. In addition to this new integrator, we implemented specific methods (regularisation integration and slow-down method) to follow the true dynamics of binaries and multiples, which is an essential element of cluster dynamics.

In this paper, we performed a first study of massive star cluster formation to benchmark the method. We focused on a single fiducial model starting from a massive turbulent molecular cloud of $10^4~\Msun$ mass and $10~\pc$ radius. Three simulations with different random initial turbulent conditions were computed with the following results:
\begin{enumerate}
    \item We found \ion{H}{II} regions --- more precisely those formed first in the simulation --- to be the main game changers for the time evolution and spatial distribution of star formation inside these clouds. Their strength and location can completely change the future of a star-forming region. We found that the formation of an \ion{H}{II} region can lead either to a very loose, globally expanding cluster in the case of efficient ionisation, or a very compact cluster when \ion{H}{II} regions are off-centred and help the gravitational focusing of the cloud.
    \item As a corollary to our first conclusion, since the early formation of \ion{H}{II} regions is a stochastic process, we found the global evolution of such a cluster to be more stochastic than when just accounting for stellar dynamics. A suite of simulations is thus needed to obtain statistically robust outcomes.
    \item We found star-formation efficiency (SFE) in our simulations to be slightly high compared to observations, with the mass conversion of gas into stars also not finished at the end of our calculations. However, the effect of our HII feedback on the SFE corroborates previous studies. Adding new physical processes, such as magnetic fields, and feedback processes including non-ionising radiation, jets, and cloud motions associated with Galactic shear should help to reduce the SFE.
\end{enumerate}

These self-consistent simulations of the formation and early evolution of embedded clusters improve previous pure N-body simulations. The aim of these is to scale up the number of simulations performed to produce statistically significant results on the dynamical evolution of clusters under the direct influence of their natal environment. Paper II will aim to characterise the star formation prescription outcomes, especially stellar mass distributions and multiplicity properties by exploring its parameter space with a larger set of simulations. This will potentially require refinements to generate a distribution of primordial binaries in the actual prescription. On the other hand, another study will be performed on high-resolution star formation simulations to connect dynamical properties seen below our sink scale to our prescription parameters.
The new methods presented here solve performance bottlenecks that have previously limited such studies. New optimisations on gas-sink interactions may further accelerate the computation. The aim is to carry out around a hundred simulations over an integration period of between $10~\Myr$ and $20~\Myr$ to have clusters with fully dispersed gas at final outputs. Ultimately, we planned to study the dynamical evolution of young clusters through statistical analysis by producing a large grid of simulations to explore a vast initial parameter space. These simulations could then be compared with observations of young clusters, including the latest data available from Gaia. 

\begin{acknowledgements}
We acknowledge M. Vaileille-Manet for his participation during the redaction of this paper. We thank the Australian-French Association for Research and Innovation (AFRAN) for financial support towards the 5th Franco-Australian Phantom users workshop held in Melbourne. DJP acknowledges Australian Research Council funding via DP220103767 and DP240103290 and thanks IPAG for hosting and financial support from both Monash and IPAG during his 2025 sabbatical in Grenoble where this paper was completed. YB thanks LabEx OSUG@2020 (Investissements d'Avenir – ANR10 LABX56) and IDEX for funding a 2-month mobility grant at Monash University. All the computations presented in this paper were performed using the GRICAD infrastructure (https://gricad.univ-grenoble-alpes.fr), which is supported by Grenoble research communities. FM, FL and IJ acknowledge funding from the European Research Council (ERC) via the ERC Synergy Grant ECOGAL (grant 855130) and from the French Agence Nationale de la Recherche (ANR) through the project COSMHIC (ANR-20-CE31-0009). We used SPLASH and Sarracen for many of the figures presented above.\citep{price_2007,Sarracen}
\end{acknowledgements}

\bibliographystyle{aa} 
\bibliography{bibliography}

\begin{appendix}
\section{Details of N-body methods}
\subsection{FSI derivation and accuracy}\label{App:FSI}
The original Leapfrog algorithm used in Phantom can be easily derived using the Hamiltonian formalism to write the equation of motion of the studied system.
This equation can be written as 
\begin{equation}
    \frac{d\mathbf{w}}{dt} = \left\{ \mathbf{w},H\right\}
\end{equation}
where $\mathbf{w}$ is the phase space vector of the system, $H$ is the system Hamiltonian and $\left\{  ,  \right\}$ is the Poisson brackets operator. The formal solution of this equation over a certain time interval $\delta t$ is then straightforwardly given by
\begin{equation}
    \mathbf{w}(t+\delta t) = e^{\delta t \, H} \mathbf{w}(t),
\end{equation}
where $e^{\delta t H}$ is an operator that evolves the system in time. All symplectic integrators come from a particular way to approximate this time evolution operator. Indeed, it can be reformulated into a product decomposition with a specific Hamiltonian that can be split into two parts, for example, a kinetic part $T$  and a potential part $U$. It gives
\begin{equation}
    e^{\delta t \, H} = e^{\delta t \, (T+U)} \approx \prod^{N_{\rm sub}}_{i=1} e^{\delta t \, c_{\mathrm{k}i} \, T}e^{\delta t \, d_{\mathrm{k}i} \, U},
    \label{sympdecomp}
\end{equation}
where $N_{\rm sub}$ corresponds to the number of subdivisions of the time evolution operator. $c_{\mathrm{k}i}$ and $d_{\mathrm{k}i}$ are coefficients (also called symplectic coefficients) that weigh the decomposition into multiple parts. The sums of these two series of coefficients need to be equal to 1 to respect the decomposition. The resulting two operator types can be associated with different steps in the time evolution. $U$ potential is always associated with a force applied on the system using $e^{\delta t U}$. $e^{\delta t U}$ is only applied on momentum coordinates
\begin{equation}
    e^{\delta t \, U} \mathbf{p}(t) = \mathbf{p}(t+\delta t).
\end{equation}
This operation is called a kick. $e^{\delta t T}$ represent the inertial part of the evolution. It will then only modify positions depending on velocities
\begin{equation}
    e^{\delta t \, T} \mathbf{x}(t) = \mathbf{x}(t+\delta t).
\end{equation}
This operation is called a drift. A system can be integrated into a given time following a succession of alternating kick and drift operations using this decomposition method. The number of kick and drift operations $N_{\rm sub}$ and a couple of coefficients $(c_{\mathrm{k}i},d_{\mathrm{k}i})$ give the integrator order of truncation. for the Leapfrog,  $N_{\rm sub} =2$  and $(c_{\mathrm{k}1},d_{\mathrm{k}1}) = (1,1/2) \; \mathrm{or} \; (1/2,1)$  and $(c_{\mathrm{k}2},d_{\mathrm{k}2}) = (0,1/2) \; \text{or} \; (1/2,0)$ depending on the order of the kick and drift operations. 
The number of operations and the symplectic coefficients can be efficiently found \citep{yoshida_construction_1990} with the Campbell-Baker-Hausdorff (CBH) formula \citep{campbell_law_1896,baker_alternants_1905}, namely
\begin{equation}
    \begin{aligned}
            \log(e^{\mathbf{X}}e^{\mathbf{Y}}) & =\mathbf{X}+\mathbf{Y}+\frac{1}{2}[\mathbf{X},\mathbf{Y}]
            \\
            +\frac{1}{12}& ([\mathbf{X},[\mathbf{X},\mathbf{Y}]] +[\mathbf{Y},[\mathbf{Y},\mathbf{X}]])+\mathcal{O}(\mathbf{X},\mathbf{Y},[\mathbf{X},\mathbf{Y}]).
    \end{aligned}
    \label{CBH}
\end{equation}
Substituting the left hand of equation (\ref{sympdecomp}) into equation (\ref{CBH}), one finds
\begin{align}
    \log(\prod^{N_{sub}}_{i=1} e^{\delta t \, c_{\mathrm{k}i} \, T}e^{\delta t \, d_{\mathrm{k}i} \, U}) &= \delta t(e_c\,\mathbf{T}+e_d\,\mathbf{U}+\frac{\delta t}{2} \, e_{cd} [\mathbf{T},\mathbf{U}] \nonumber \\
    &+\frac{\delta t}{12} (e_{ccd}[\mathbf{T},[\mathbf{T},\mathbf{U}]] +e_{dcd}[\mathbf{U},[\mathbf{T},\mathbf{U}]]) \nonumber \\
    &+\mathcal{O}(\mathbf{T},\mathbf{U},[\mathbf{T},\mathbf{U}]),
    \label{CBHsymp}
\end{align}
where all $e$ prefactors are called error coefficients and are functions of $(c_{\mathrm{k}i},d_{\mathrm{k}i})$. The first two coefficients correspond to the sum of $c_{\mathrm{k}i}$ and $d_{\mathrm{k}i}$, which by definition needs to equal unity. Thus, $e_c$ and $e_d$ are equal to unity. One can refactor equation (\ref{CBHsymp}) using these equalities according to
\begin{equation}
    \log\left(\prod^{N_{\rm sub}}_{i=1} e^{\delta t \, c_{\mathrm{k}i} \, T}e^{\delta t \, d_{\mathrm{k}i} \, U}\right) = \delta t\left[H+H_{\rm err}(\delta t)\right] = \delta t  H_{\rm neigh}.
    \label{neighH}
\end{equation}

This new form hints at what is needed for a symplectic integration. What is evolved in such a scheme is not the true Hamiltonian $H$ but a neighbour of this one $H_{\rm neigh}$ which differs from an error called $H_{\rm err}$. The goal of constructing high-order symplectic integrators is then to minimise this error quantity by setting the next error coefficients to zero. However, it has been proven \citep{sheng_solving_1989,suzuki_general_1991} that after second order $(c_{\mathrm{k}i},d_{\mathrm{k}i})$ coefficients must contain some negative values to set the next order coefficient to zero. However, negative time steps are problematic when integrating dissipative systems that are common in astrophysics. The solution proposed by \cite{suzuki_general_1991} was to add one of the first terms of the error Hamiltonian produced by a Leapfrog scheme into the true one. One can then erase all error terms of the fourth order. Nonetheless, the part added to the classical Hamiltonian still needs to be computed. It is possible to write the new Hamiltonian as 
\begin{equation}
    H = T+U+\frac{1}{48} \delta t^2[U,[T,U]].
\end{equation}
In this equation, $[U,[T,U]]$ is a composition of two commutators between $U$ and $T$. \cite{dehnen_symplectic_2017} showed that this commutator corresponds to an extra potential quantity in the Hamiltonian defined as 
\begin{equation}
[U,[T,U]] = -\sum_{i}^N \frac{1}{m_i}\left(\frac{\partial U}{\partial r_i}\right)^2 = -\sum_{i}^N m_i \lVert \mathbf{a}_i \rVert.
\end{equation}

\begin{figure*}
     \centering
     \begin{subfigure}[b]{0.49\textwidth}
         \centering
         \includegraphics[width=\textwidth]{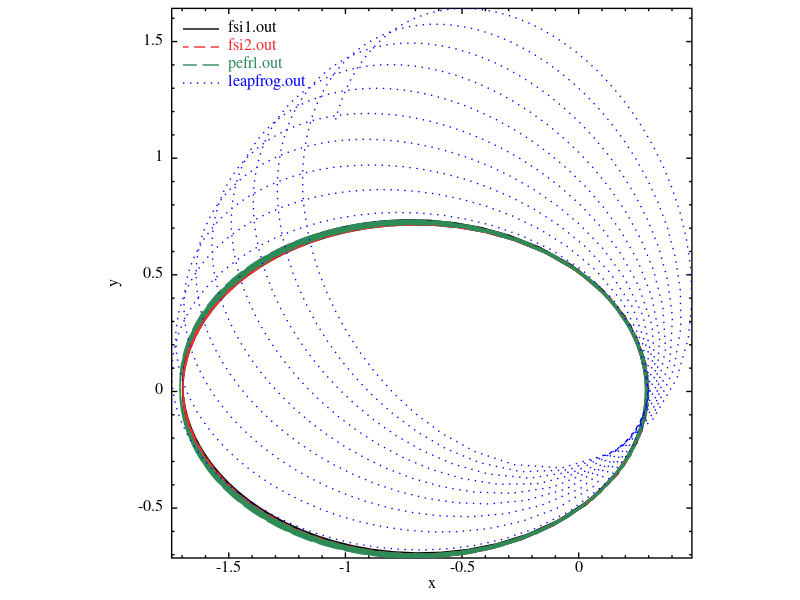}
         \caption{\small 10 orbits of a point mass around a fixed potential. Initial eccentricity is set to 0.7 and the semi-major axis is 1. The amount of (spurious) orbital precession can be seen to inversely correlate with integrator precision.}
         \label{fig:orbit_int}
     \end{subfigure}
     \begin{subfigure}[b]{0.49\textwidth}
         \centering
         \includegraphics[width=\textwidth]{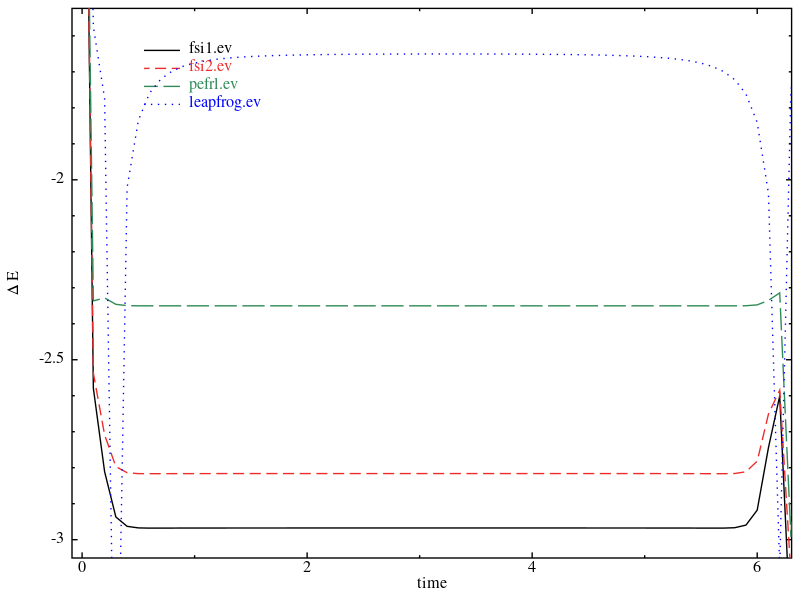}
         \caption{Logarithm of the relative energy of the particle on 1 orbit.}
         \label{fig:DE_int}
     \end{subfigure}
     \caption{Evolution of a point mass around a fixed potential}
     \label{fig:int_precision}
\end{figure*}

Figure \ref{fig:orbit_int} and \ref{fig:DE_int} illustrate the superior accuracy of the FSI method by showing the trajectory of one point mass around a fixed gravitational potential during 10 orbits and its energy conservation (logarithm of the relative energy) for one orbit, respectively. FSI1 and FSI2 correspond, respectively, to the original gradient force calculation and the extrapolation method. The initial eccentricity of the body is large ($e=0.7$) to show the orbital precession that can be introduced by numerical error. A large timestep is also deliberately chosen (around 60 steps per orbit) to amplify numerical imprecision. As expected, energy conservation and orbit precession are much improved with a fourth-order scheme than with a simple leapfrog scheme. It also confirms the \cite{chin_forward_2005} result that forward schemes have better conservation properties. 

\subsection{Few body problem regularised by TTL} \label{APP:TTL}
Let's introduce a general N-body (only Newtonian) system that we want to regularise with the Time Transformed Leapfrog (TTL) scheme. The equations of motion are given by
\begin{equation}
    \ddot{\mathbf{r}_{i}} = G\sum^N_{i = 1}\sum^N_{i\neq j}m_j\frac{\mathbf{r}_j-\mathbf{r}_j}{r_{ij}^3} = {\bf F}_i
\end{equation}
where $N$ is the number of bodies and $G$ the gravitational constant and $r_{ij}$ is the distance between body $i$ and $j$. For convenience $G=1$. Now, let's introduce the time transformation
\begin{equation}
    \mathrm{d}s = \Omega(\mathbf{r})\mathrm{d}t,\label{ttrans}
\end{equation}
where $\Omega(\mathbf{r})$ is an arbitrary time transformation function. To follow what is done in \cite{mikkola_time-transformed_2002} and \cite{wang_slow-down_2020}, this function can be defined as
\begin{equation}
    \Omega(\mathbf{r}) = \sum^N_{i = 1}\sum^N_{i\neq j}\frac{m_i m_j}{r_{ij}}.
\end{equation}
One can now set a new auxiliary variable, often called $W=\Omega$, to rewrite the system according to
\begin{eqnarray}
    \frac{\mathrm{d}r_i}{\mathrm{d}s}&=&\frac{v_i}{W}\\
    \frac{\mathrm{d}v_i}{\mathrm{d}s}&=&\frac{F_i}{\Omega}\\
    \frac{\mathrm{d}t}{\mathrm{d}s}  &=&\frac{1}{W};
\end{eqnarray}
where index $i$ indicates one body of the system. The addition of a new variable gives a supplementary variable in the system that can be extracted from the differential equations.
\begin{equation}
    \frac{\mathrm{d}W}{\mathrm{d}t} = \sum_{i=1}^N \frac{\partial \mathbf{r}_i}{\partial t}\cdot \frac{\partial \Omega}{\partial \mathbf{r}_i},
\end{equation}
which can be expressed using our new ``time'' coordinate
\begin{equation}
    \frac{\mathrm{d}W}{\mathrm{d}s} = \sum_{i=1}^N \frac{1}{\Omega}\frac{\partial \mathbf{r}_i}{\partial t}\cdot \frac{\partial \Omega}{\partial \mathbf{r}_i}.
\end{equation}
This new variable $W$ acts like an associated momentum for the time, which means that it will have the same behaviour as a velocity in a symplectic scheme. With that in mind, it is then possible to form one drift and one kick operation from this system, given respectively by
\begin{eqnarray}
    \mathbf{r}_{i*} &=& \mathbf{r}_i + c_k\delta t\frac{\mathbf{v}_i}{W} \\
    t_*  &=& t + \frac{c_k\delta t}{w},
\end{eqnarray}
where $c_k$ is the drift coefficient and $\delta t$ is the time step. Then
\begin{eqnarray}
   \mathbf{v}_{i*/2} &=&  \frac{d_k\delta t}{2}\frac{\mathbf{F}_i(\mathbf{r}_{i*})}{\Omega(\mathbf{r}_{i*})} \\
   W_* &=& d_k\delta t \sum_{i=1}^N \frac{\mathbf{v}_{i*/2}}{\Omega(\mathbf{r}_{i*})}\cdot \frac{\partial\Omega(\mathbf{r}_{i*})}{\partial\mathbf{r}_{i*}}\\
    \mathbf{v}_{i*} &=&  \frac{d_k\delta t}{2}\frac{\mathbf{F}_i(\mathbf{r}_{i*})}{\Omega(\mathbf{r}_{i*})}
\end{eqnarray}
where $d_k$ is the kick coefficient and $*$ represents updated variables. As $W$ depends on the velocities of each body, it is needed to update both variables simultaneously. To achieve that, the velocity kick is split in half with a $W$ kick between the two parts. These two operations can be successively applied to the system to integrate its motion using any kind of symplectic integrator. 

\subsection{Discussion on slow-down factor} \label{APP:SD}
As has been shown above, $\kappa$ quantifies the number of original periods where the perturbation applied to the binary can be averaged. On a highly perturbed binary, it is impossible to average the perturbation on multiple periods, which means $\kappa = 1$. On the other side, a hard binary can stay unperturbed during a whole simulation. The averaging is not risky at all, and then $\kappa$ can stay at a high value. That being said, one can see that this factor needs to scale with the strength of the perturbation. The solution proposed originally by \cite{mikkola_slow-down_1996} is to compute the relative tidal perturbation at the maximum apocenter ( $2a$ with $e=1$), giving
\begin{equation}
    \gamma = \frac{8(m_p+m_s)a^3}{m_pm_s}\sum_{j=1,j\neq p,s}^N\frac{m_j}{r_{bj}^3},
\end{equation}
where $a$ is the semi major axis of the binary,$p$ and $s$ point on the binary components. $r_{bj}$ is the distance between any perturbers and the binary centre of mass. $\kappa$ value is then equal to
\begin{equation}
    \kappa = \frac{\gamma_0}{\gamma},
\end{equation}
where $\gamma_0$ is a safety factor to adjust the precision wanted by the user, often chosen between $10^5$ and $10^6$. \cite{wang_slow-down_2020} chose something slightly different for this quantity. In their work, they proposed
\begin{equation}
    \kappa = \frac{m_pm_s}{m_p+m_s(a(1+e))}\sum_{j=1,j\neq p,s}^N\frac{r_{bj}^3}{m_j}.
\end{equation}
Instead of using the maximum possible apocenter of the binary, they chose the actual apocenter. The most different part of their proposition is the summation. They directly sum the cubed distances scaled by the mass of each perturber. Even if these two expressions look similar, they have totally different behaviour. For example, let's consider a system composed of one hard binary that can be slowed down and two single perturbers. To simplify the system, each bodies have the same mass. One perturber is far from the binary, producing weak perturbations. The other one is on a hyperbolic trajectory with the binary close to its periapsis. It then produces a strong perturbation. In theory, this scenario should produce a slowdown factor near unity due to the close encounter with perturber two. The first expression will first sum the inverse cube distance, which will give a large number (since the strong perturber dominates the summation). Taking the inverse of this summation gives a number less than unity (which means no slowdown in this framework). The first expression has good behaviour, but with the second expression, the dominant perturber is the weaker one, which will give a high $\kappa$ value. This expression is then potentially wrong. Therefore, in this work, we stick with the original expression, which leads to more robust behaviour by changing only the apocenter expression.

Finally, the $\kappa$ value of each binary in the system needs to be updated regarding the perturbation modification in the system. However, this evolution needs to be slow to conserve the secular evolution properties. \cite{wang_slow-down_2020} proposed a limit value for $\kappa$ that is controlled by the changing rate of the perturbation. Formally, it is written as
\begin{equation}
    \kappa_{\rm max} = \frac{c\|\langle \mathbf{r}_{bj}\rangle_m \|}{P\|\langle \mathbf{v}_{bj}\rangle_m \|}\label{kappamax}
\end{equation}
where $c=0.1$ is a safety factor, $P$ is the period of the binary, $\mathbf{v}_{bj}$ is the relative velocity vector between perturbers and the binary CoM. $\langle\rangle_m$ refers to a mass-weighted average.

\section{HII region unit test} \label{App:HII}
\begin{figure}
    \centering
    \includegraphics[width=0.5\textwidth]{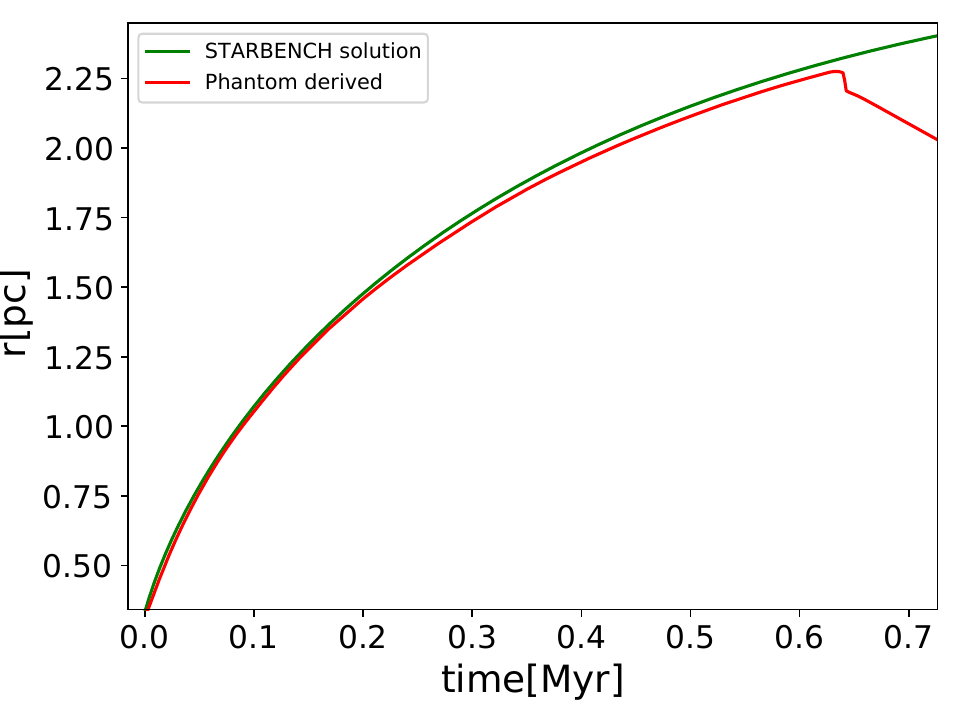}
    \caption{Time evolution of the HII region front radius. The green line is the analytical solution found in \cite{Bisbas_2015}. The red line is the Phantom-derived solution. The sudden drop after $0.6\,\Myr$ is due to the front shock reflection onto the boundary particles.}
    \label{fig:HIIR}
\end{figure}

\begin{figure}
    \centering
    \includegraphics[width=0.5\textwidth]{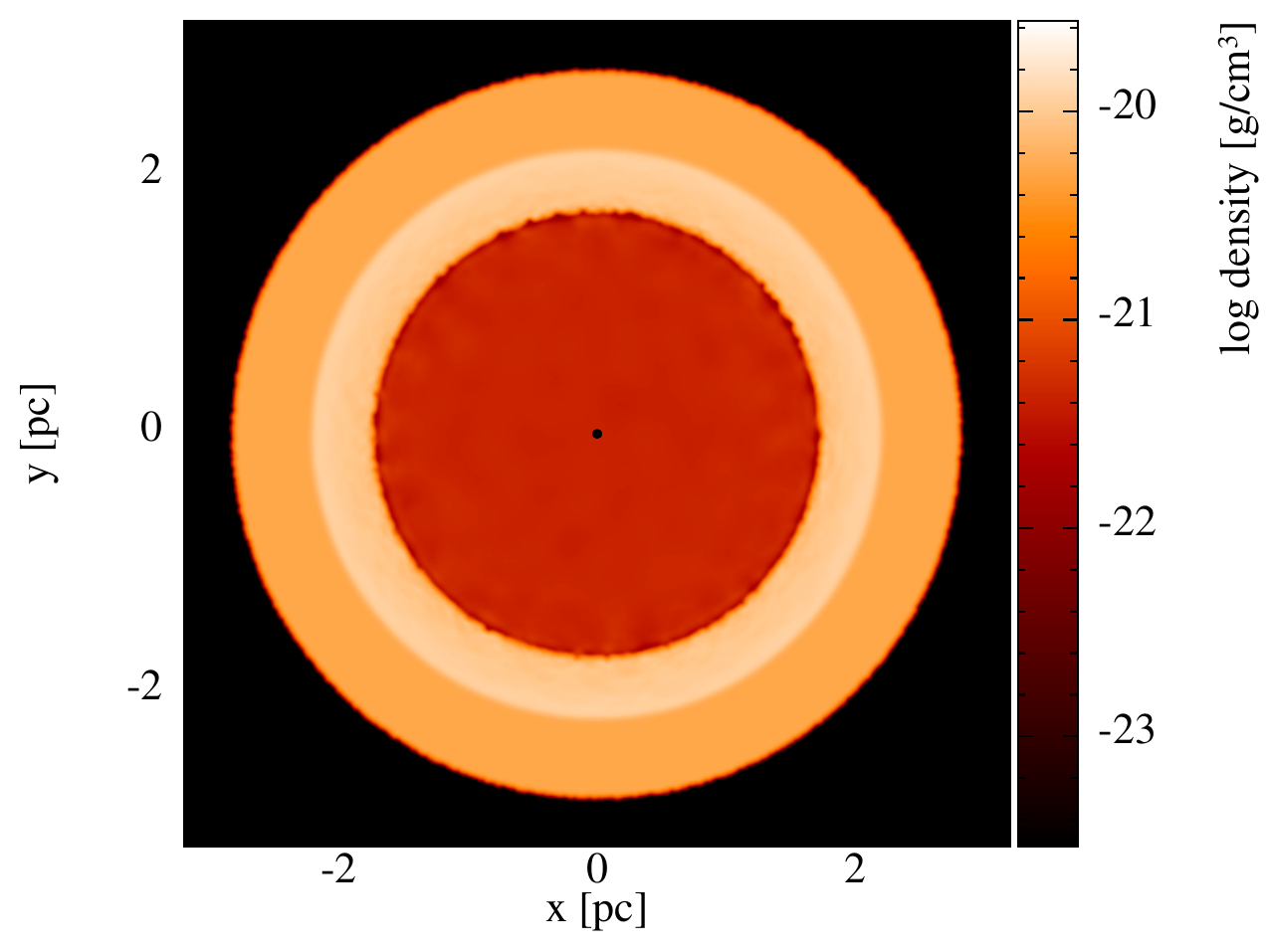}
    \caption{Density slice of a uniform density sphere where a HII region expands from its centre. A central stellar object produces many ionising photons that heat the gas around it. Two media (hot and cold) are separated by a front shock that propagates inside the cold sphere, sweeping up the gas. A denser shell of material forms just next to the ionisation front. }
    \label{fig:HII}
\end{figure}

We performed a simple test in order to validate the HII feedback implemented into this work, following what was proposed in \cite{Bisbas_2015}. They described two main setups to check the validity of HII region expansion at two different time scales (early and late) compared to analytical solutions. Such solutions exist only in very specific and simple systems. The selected test (late) tries to expand an HII region produced by a $40 \Msun$ star in the centre of a homogeneous sphere of hydrogen atomic gas of $2.9 \pc$ in radius. Here, we set the cloud mass to $8\times10³\Msun$, which was divided into 8 million SPH particles. The neutral temperature was set to $1000 \, \mathrm{K}$. Boundary particles are also mandatory at the edge of the cloud to avoid relaxing into empty space, which could lead to an artificial acceleration of the HII bubble. Figure~\ref{fig:HIIR} plots the HII region radius derived by Phantom, while Figure~\ref{fig:HII} shows a density slice of the cloud.
Our implementation agrees with the \cite{Bisbas_2015} analytical solution during $0.6 \Myr$ and comes close to the stagnation radius equal to $2.31 \,\pc$. The sudden drop after $0.6 \Myr$ is due to the shock front reflecting from boundary particles which compresses the HII bubble, forcing it to shrink suddenly. This phenomenon is not present in \cite{Bisbas_2015} work. Nothing is detailed on boundary conditions for SPH runs. Nonetheless, the satisfactory agreement on the first part is sufficient to validate the implementation, as boundary conditions in large clouds have already been tackled in \ref{sec:HII}.

\section{SDAR unit test} \label{App:SDARtest}

We performed a unit test taken from \cite{wang_slow-down_2020} to validate our SDAR implementation. This test consists of a hierarchical triple system (binary-single) with a hard eccentric inner binary orbiting another single star. The outer orbit is eccentric as well, with a semi-major axis ratio of $1000$ between the inner and outer binary. All orbital parameters are summarised in Table 1  of \cite{wang_slow-down_2020}.

\begin{figure}
    \centering
    \includegraphics[width=\linewidth]{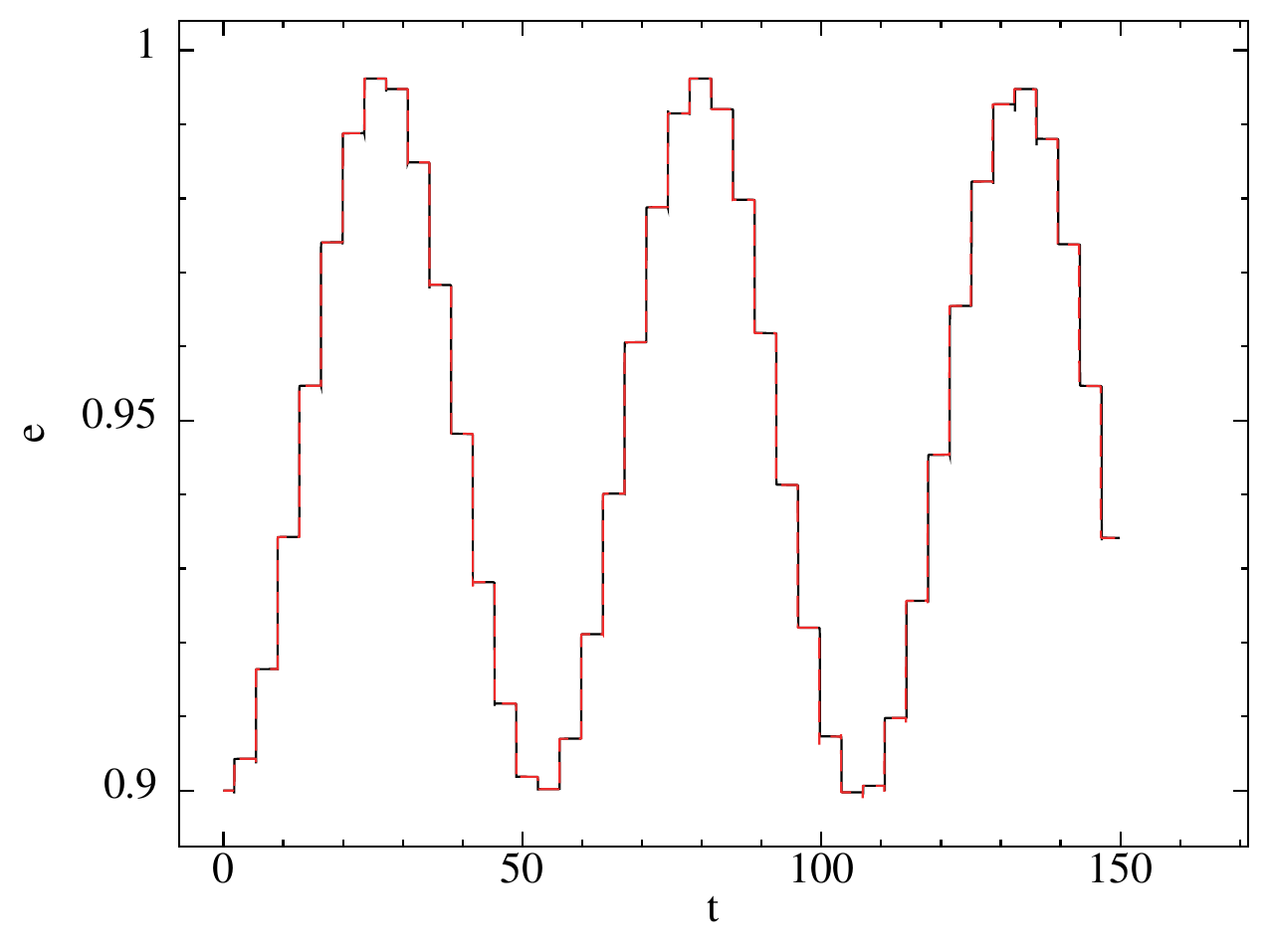}
    \caption{Time evolution of the inner binary eccentricity. This eccentricity oscillates under the perturbation of the single star orbiting the binary in a very eccentric orbit.}
    \label{fig:Eccentricity}
\end{figure}

\begin{figure}
    \centering
    \includegraphics[width=\linewidth]{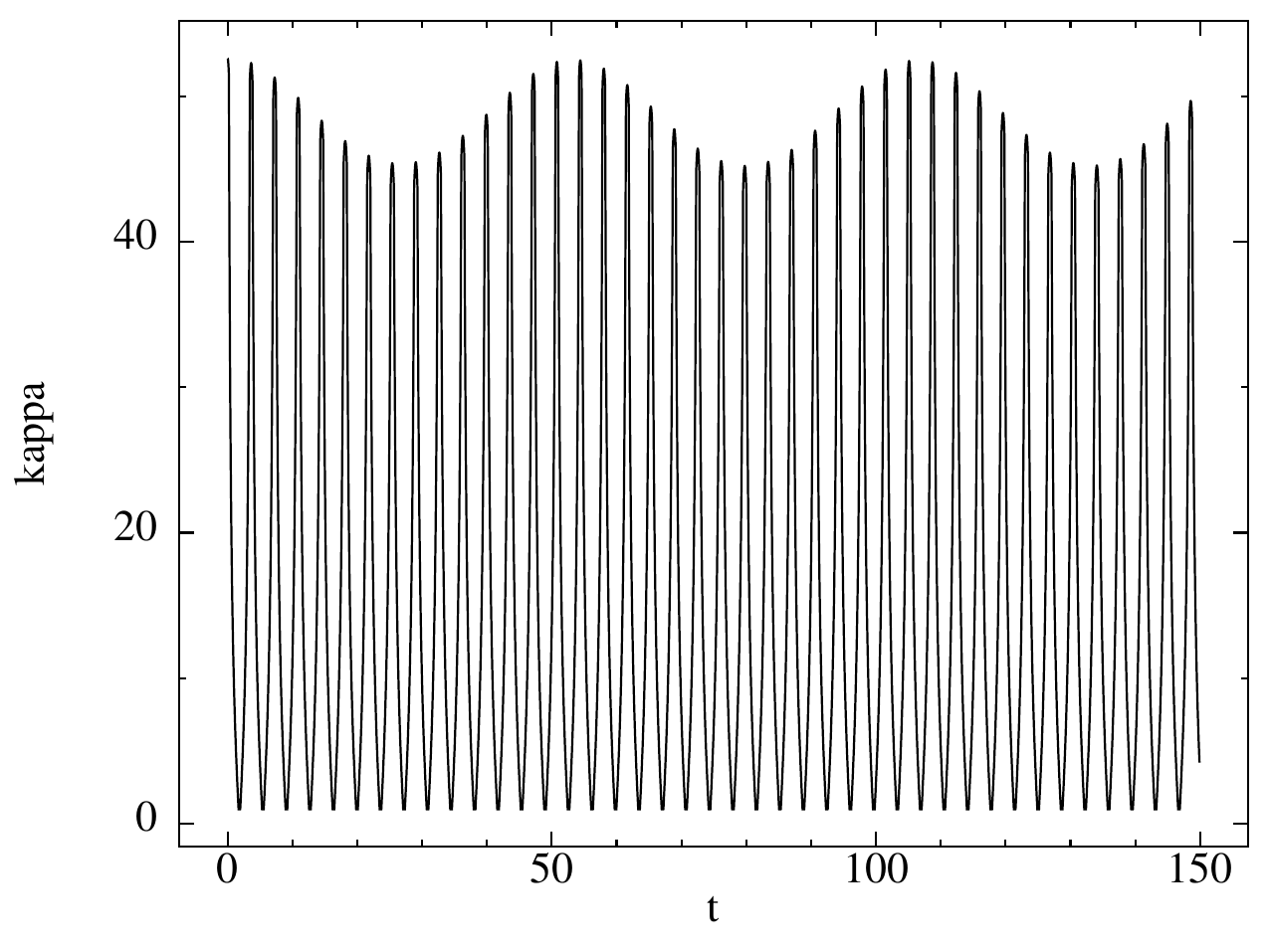}
    \caption{Time evolution of the slowdown factor $\kappa$. This one oscillates on a shorter timescale that corresponds to the outer binary period. When the single star reaches the apoapsis of its orbit. $\kappa$ reaches its maximum at this moment. Oppositely and as expected, $\kappa$ decreases during its descent to the periapsis. At this moment, slowing down the inner binary motion is not possible, therefore $\kappa = 1$}
    \label{fig:kappa}
\end{figure}

Kozai-Lidov oscillations \citep{kozai1962,lidov1962}) are expected on the inner eccentricity and inclination in such systems as the outer star will perturb the inner binary at each periapsis. Such a system is a good test to verify if our SDAR module can correctly model this oscillation with and without the slowdown method. \Cref{fig:Eccentricity} shows the eccentricity of the inner binary over time. As expected, A clear oscillation between $0.9$ and $1.$ is visible, and both methods converge on the same solution. \Cref{fig:kappa} shows the evolution of $\kappa$ slow-down factor over time. This one oscillates between around $50$ and $1$ depending on the outer binary orbital phase. If the single outer star is at apoapsis, its perturbation on the inner binary is minimal; therefore, $\kappa$ is high. Its value decreases accordingly down to $1$ as the star reaches the periapsis of the outer orbit. In this configuration, the perturbation is maximal. Hence, slowing down the inner binary is not possible. These results also converge with what was found in \cite{wang_slow-down_2020}.

\end{appendix}

\end{document}